\documentclass[12pt]{article}
\usepackage{graphicx}
\usepackage{amsmath}
\usepackage{amsfonts}
\usepackage{amssymb}

\begin{document}

\author{Arkady L.Kholodenko\thanks{375 H.L.Hunter Laboratories ,Clemson
University,Clemson ,SC 29634-1905 .USA .e-mail address :string@ clemson.edu}}
\title{Use of Quadratic Differentials for Description of Defects and Textures in
Liquid Crystals and 2+1 Gravity}
\date{}
\maketitle
\begin{abstract}
The theory of measured foliations which is discussed in Part I in connection
with the train tracks and meanders is shown to be related to the theory of
Jenkins-Strebel quadratic differentials by Hubbard and Masur (Acta
Math.\textbf{142 },221 (1979)).Usefulness of the measured foliations for
description of defects and textures in liquid crystals had been recognized by
Poenaru and Langevin some time ago.Based on the theory of quadratic
differentials , it is demonstrated that this formalism not only provides the
adequate description of defects and textures in liquid crystals but also is
idealy suited for study of 2+1 classical gravity which was initiated in the
seminal paper by Deser , Jackiw and 't Hooft (Ann.Phys.\textbf{152},220
(1984)).Not only their results are reproduced but,in addition,many new results
are obtained.In particular,using the results of Rivin (Ann.Math.\textbf{139}%
,553 (1994)) the restriction on the total mass of the 2+1 Universe is removed
.It is shown ,that the masses can have only discrete values and,moreover, the
theoretically obtained sum rules forbid the existence of some of these values.
The dynamics of 2+1 gravity which is associated with the dynamics of train
tracks(Part I) , is reinterpreted in terms of the emerging hyperbolic
3-manifolds .The existence of knots and links associated with complements of
these 3-manifolds is highly nontrivial and requires careful proofs.The paper
provides a concise introduction into this topic.The discussion of some
connections with related physical problems which could be treated with help of
quadratic differentials is also provided.These include (but not limited
to):string theory,classical and quantum billiards,dynamics of fracture
,statics and dynamics of dislocations and disclinations in solids,etc.
\end{abstract}

\section{Introduction}

In the previous paper[$1]$ (Part I ) we have discussed some aspects of the
theory of measured foliations for 2 dimensional surfaces without providing
sufficient mathematical proofs.In this part we would like to provide a
self-contained mathematical justification of the obtained results.Our
presentation is deliberately pedagogical enough to be accessible not only to
the experts on gravity but to the interested readers in the areas of condensed
matter physics.

The theory of measured foliations is mathematically connected with the theory
of quadratic differentials as was demonstrated by Hubbard and Masur$[2]$.
\ Therefore ,we mainly will concentrate our efforts on this connection in this
part of our paper. The usefulness of the measured foliations for the
description of defects and textures in liquid crystals was recognized for some
time$[3,4]$ .More recently,the usefulness of quadratic differentials for
description of bosonic strings was recognized in Ref.[5].Additional nontrivial
results for strings which involve quadratic differentials were obtained in
Ref.[6]Here ,we argue that the quadratic differentials not only naturally
occur in 2+1 gravity but, in the Discussion section, we indicate that they may
also be helpful in the theories of motion in classical and quantum
billiards,theories of elasticity and dynamics of fracture ,etc .In section 2
we provide an auxiliary introduction to the isoperimetric
inequalities.Although the existing monographs on quadratic differentials do
not contain this information collected in a special section$[7,8]$ ,they
,nevertheless, assume that the reader is familiar with it.We found,that for
the nonexperts,the isoperimetric inequalities provide the most natural
background necessary for understanding of the physical meaning of quadratic
differentials.Although most of the results presented in this section are
known,they are scattered in the literature and this circumstance,we feel, is
sufficient for presentation of them in this paper.In section 3 these results
acquire new meaning when we discuss essentials of quadratic differentials. In
the mathematical physics literature there is already an excellent treatment of
this topic,e.g.see Ref.[5] ,nevertheless,we feel that our presentation could
be considered as complementary.Our discussion in this section is subordinated
to our intention to use the quadratic differentials in the theory of liquid
crystals and gravity. Applications to liquid crystals are discussed in section
4.The results of this section are aimed to explain the physical meaning of
quadratic differentials in terms of known (to physicists)results from the
differential geometry of surfaces with singularities.We also provide some
proofs in support of the results discussed in section 5 (Part I) .In this
section the surface energy , Eq.(5.3), was used in calculations without
mathematical justification. Next,in section 5 of \textbf{this} part of our
work we discuss applications of quadratic differentials to 2+1 gravity.The
presentation of this section is strongly influenced by the seminal work ,
Ref.[9] ,by Deser , Jackiw and 't Hooft. We reanalyze and extend their results
using known in mathematical literature connection between the quadratic
differentials and conical (surface) singularities $[10]$ .We reobtain this
connection in the way somewhat different from that given in Ref.[10] and
then,we extend this connection with help of recently obtained more
comprehensive mathematical results by Rivin$[11]$ .This allows us to remove
the restriction on the maximal total mass of 2+1 Universe which follows from
the results of Ref.9 .We also demonstrate that 2+1 gravity to some extent can
be considered as a special case of the Poincare-Hopf index theorem (Eq.(1.9)
,Part I )and using the results of Hopf $[12]$ we argue that the masses in such
theory should take only discrete values.That is, already at the classical
level,we obtain a sort of quantization condition for masses.The results of
this section are elaborated in section 6 where more advanced topics are
briefly discussed .In particular,we argue that not only masses should be
quantized but , in addition , the obtained mass spectrum is subject to some
selection rules which forbid existence of particles with certain values of
quantized masses.This result follows directly from the theory of quadratic
differentials$[13]$ .We also briefly study the effects of inclusion of the
cosmological term into Einstein's equations of 2+1 gravity at the classical
level.This inclusion is studied from both mathematical and physical points of
view.In particular,the connection with the nonperturbative treatment of
bosonic string theory by Takhtadjian$[14]$ is briefly discussed.Since in Part
I (section 4) we had developed theory of measured foliations based on the
train tracks ,we provide in section 6 some connections of this theory and the
theory of Teichm\"{u}ller spaces in order to clarify the physical processes
which are responsible for the topology nonpreserving moves for train tracks
which are depicted in Fig. 24(Part I) .These processes are responsible for
phase changes (reducible,periodic, pseudo-Anosov)discussed in Part I.
Finaly,we briefly discuss how the motion in Teichm\"{u}ller space is related
to the evolution in the Minkovski spacetime.This treatment is complementary to
that developed by Moncrief$[15]$ and serves to sketch the connections between
2+1 gravity ,hyperbolic 3-manifolds and the theory of knots and links(for some
additional gravity-nonrelated physical applications of the theory of knots and
links,please,consult Ref.[16]).Some important auxiliary results related to
knots/links and the hyperbolic 3-manifolds are presented in the Appendix.This
is motivated by the fact ,that from the mathematical standpoint,the existence
of knots and links in 2+1 gravity is by no means selfobvious.In the main text
we provide sufficient arguments in order to demonstrate the existing
difficulties .In the Appendix we provide some sketch of very recent
mathematical results in support of the existence of knots and links in 2+1
gravity. Our exposition of mathematical results related to hyperbolic
3-manifolds is rather terse(unlike other subjects which we treat with
sufficient details).Nevertheless,it is included in this paper since it is
logically connected with the rest .We plan to return to these more advanced
subjects in the future publications.

Additional physical problems which may be treated by the methods developed in
Parts I and II are briefly listed in the Discussion section.These problems
include(but not restricted to): classical and quantum billiards,dynamics of
fracture ,dislocations in solids,Temperley-Lieb algebra related to meanders
and its connection with the invariants of 3-manifolds,etc.

\section{Some important isoperimetric inequalities}

Consider a closed simple (without self-intersections) planar curve
$\mathcal{C}$ of length L and let A be the area which is enclosed by
$\mathcal{C}$ ,then%

\begin{equation}
\text{L}^{2}\geq4\pi\text{A ,} \tag{2.1}%
\end{equation}
where the equality holds only for a circle.Physical applications of this
inequality had been recently discussed in Ref.[17].Here ,we are mainly
concerned with the methods of proving the inequality (2.1).

Since analytically the length L is given by%

\begin{equation}
\text{L=}\int\limits_{a}^{b}d\tau\sqrt{\left(  \frac{dx}{d\tau}\right)
^{2}+\left(  \frac{dy}{d\tau}\right)  ^{2}} \tag{2.2}%
\end{equation}
and the area A is known to be given by%

\begin{equation}
\text{A=-}\int\limits_{a}^{b}d\tau y\frac{dx}{d\tau} \tag{2.3}%
\end{equation}
,we can use as imple identity (obtained with help of parametrization:$\tau
=\frac{2\pi}{\text{L}}s$ )%

\begin{equation}
\int\limits_{0}^{2\pi}d\tau\left[  \left(  \frac{dx}{d\tau}\right)
^{2}+\left(  \frac{dy}{d\tau}\right)  ^{2}\right]  =\int\limits_{0}^{2\pi
}d\tau\left(  \frac{ds}{d\tau}\right)  ^{2}=\frac{\text{L}^{2}}{2\pi}\text{ }
\tag{2.4}%
\end{equation}
in order to obtain%

\[
\text{L}^{2}-4\pi\text{A=2}\pi\int\limits_{0}^{2\pi}d\tau\left[  \left(
\frac{dx}{d\tau}\right)  ^{2}+\left(  \frac{dy}{d\tau}\right)  ^{2}%
+2y\frac{dx}{d\tau}\right]  \text{ .}
\]%

\begin{equation}
=2\pi\int\limits_{0}^{2\pi}d\tau\left(  \left(  \frac{dx}{d\tau}\right)
+y\right)  ^{2}+2\pi\int\limits_{0}^{2\pi}d\tau\left[  \left(  \frac{dy}%
{d\tau}\right)  ^{2}-y^{2}\right]  \tag{2.5}%
\end{equation}
Hence,to prove the inequality (2.1) we have to prove that%

\begin{equation}
\int\limits_{0}^{2\pi}d\tau\left(  \frac{dy}{d\tau}\right)  ^{2}\geq
\int\limits_{0}^{2\pi}d\tau y^{2} \tag{2.6}%
\end{equation}
This,however,may or \textbf{may not }be the case,e.g. if y=const the above
inequality certainly fails.It is rather easy to prove $[18]$ that

\textbf{Theorem} \textbf{2.1}.\textit{If }$y$\textit{(}$\tau)$\textit{ is a
smooth function with period 2}$\pi$\textit{ and if }$%
{\textstyle\int\limits_{0}^{2\pi}}
d\tau y(\tau)=0,$\textit{ then the inequality (2.6) holds and becomes an
equality if and only if }$y(\tau)=a\cos\tau+b\sin\tau$\textit{ ,i.e. when the
trajectory is a circle(physical applications of this fact are discussed in Ref.[17])}

The above result (2.1) can be extended for the case when the curve may have
self-intersections[19]. We are not going to need this case however.Instead,we
shall consider the extension of this result to the nonflat surfaces. In the
case of a sphere of radius R ,it could be shown $[20]$ that%

\begin{equation}
\text{L}^{2}\geq4\pi\text{A - }\frac{\text{A}^{2}}{\text{R}^{2}} \tag{2.7}%
\end{equation}
with equality holding only for a circle on the sphere.In the case of a
pseudosphere (surface of constant negative curvature )of radius R= i ,
i=$\sqrt{-1}$ ,we obtain, instead of (2.7),the following result:%

\begin{equation}
\frac{\text{L}^{2}}{\text{A}}\geq\text{A + 4}\pi\text{ .} \tag{2.8}%
\end{equation}
Obviously,if (2.8) holds ,then the inequality of the type given by (2.1) holds
as well.Beacause of this, following Ahlfors $[21]$ ,let us define the extremal
length $\lambda(\Gamma)$ via%

\begin{equation}
\lambda(\Gamma)=\sup_{\rho}\frac{L^{2}(\rho)}{A(\rho)} \tag{2.9}%
\end{equation}
where $\Gamma$ is some closed set of curves and $\rho$ is the metric of the
surface defined in such a way that
\begin{equation}
L_{C}\left(  \rho\right)  =\int\limits_{C}\rho\left|  dz\right|  \tag{2.10}%
\end{equation}
and%

\begin{equation}
A_{\Delta}(\rho)=\iint\limits_{\Delta}\rho^{2}dxdy \tag{2.11}%
\end{equation}
where $\Delta$ is the area enclosed by the contour C$\in\Gamma$ and $z=x+iy$ .

\textbf{Remark} \textbf{2.2}. \textit{The above definition of }$\lambda
(\Gamma)$\textit{ can be extended to open curves as well if we have surfaces
with boundaries.}

To get a feeling of the above results ,let us consider a rectangle \^{R} with
sides $a$ and $b$ and let C be some curve which joins the opposite sides of
the rectangle. Then,for any $\rho$%

\begin{equation}
\int\limits_{0}^{a}\rho\left(  x+iy\right)  dx\geq L_{C}(\rho) \tag{2.12}%
\end{equation}
and,from here,
\begin{equation}
\int\limits_{0}^{b}\int\limits_{0}^{a}\rho dxdy\geq bL_{C}(\rho)\text{ .}
\tag{2.13}%
\end{equation}
With help of (2.11) and (2.13) we obtain as well
\begin{equation}
b^{2}L^{2}(\rho)\leq ab\int\limits_{0}^{a}\int\limits_{0}^{b}\rho
^{2}dxdy=abA_{\hat{R}}(\rho)\text{ .} \tag{2.14}%
\end{equation}
The above result was obtained with help of the Schwarz-type inequality $[22]$
\begin{equation}
\left[  \iint\limits_{\hat{R}}\rho\cdot1dxdy\right]  ^{2}\leq\iint
\limits_{\hat{R}}\rho^{2}dxdy\iint\limits_{\hat{R}}dxdy\cdot1=ab\iint
\limits_{\hat{R}}\rho^{2}dxdy\text{ \quad\quad.} \tag{2.15}%
\end{equation}
The inequality (2.14) produces ,in turn,
\begin{equation}
\frac{a}{b}\geq\frac{L^{2}(\rho)}{A_{\hat{R}}(\rho)}\equiv\lambda\left(
\Gamma\right)  \text{ \quad.} \tag{2.16}%
\end{equation}
At the same time,if we choose $\rho=1$ inside \^{R} (and $\rho=0$
outside),then $L(\rho)=a$ and $ab=A(\rho)$ so that $\lambda\left(
\Gamma\right)  \geq\frac{a}{b}.$ From here we arrive at the conclusion ,that
for the rectangle $\hat{R}$%
\begin{equation}
\frac{a}{b}=\lambda(\hat{R})\text{ .} \tag{2.17}%
\end{equation}
By definition,the \textbf{modulus }$M=\frac{b}{a}=\lambda^{-1}$ or ,in
general,
\begin{equation}
M(\Gamma)=\inf_{\rho}\frac{A(\rho)}{L^{2}(\rho)}\text{ \quad.} \tag{2.18}%
\end{equation}
With help of M we obtain for the rectangle $\ \hat{R}$ the following result :
\begin{equation}
A_{\hat{R}}(\rho)=\iint\limits_{\hat{R}}\rho^{2}dxdy\geq a^{2}M\text{ \quad.}
\tag{2.19}%
\end{equation}
This happens to be central result for the entire development as we shall
demonstrate below in the rest of this work.To this purpose ,let us discuss the
related problem about the calculation of the modulus M for the annulus
$\check{D}$ ,i.e. doubly connected region made of two concentric rings $C_{1}$
and $C_{2}$ of radius $r_{1}<r_{2}$ .By choosing the polar system of
coordinates,we obtain ,by analogy with (2.12),
\begin{equation}
L(\rho)\leq\int\limits_{0}^{2\pi}\rho\left(  re^{i\varphi}\right)
rd\varphi\tag{2.20}%
\end{equation}
for any closed curve \textbf{which separates }$C_{1}$ \textbf{and} $C_{2}$
.From here,
\[
\frac{L(\rho)}{r}\leq\int\limits_{0}^{2\pi}\rho d\varphi
\]
and
\begin{equation}
L(\rho)\ln\left(  \frac{r_{2}}{r_{1}}\right)  \leq\iint\limits_{\check{D}}\rho
drd\varphi\text{ .} \tag{2.21}%
\end{equation}
This resembles very much (2.13) and,therefore,by analogy with (2.14) we obtain
,
\begin{equation}
L^{2}(\rho)\ln^{2}\left(  \frac{r_{2}}{r_{1}}\right)  \leq\left[
\iint\limits_{\check{D}}\rho drd\varphi\right]  ^{2}\text{ .} \tag{2.22}%
\end{equation}
Again,using the Schwarz inequality,
\begin{equation}
\left[  \iint\limits_{\check{D}}\frac{1}{r}\rho rdrd\varphi\right]  ^{2}%
\leq\left[  \int\limits_{r_{1}}^{r_{2}^{{}}}dr\int\limits_{0}^{2\pi}\frac
{dr}{r}d\varphi\right]  \cdot\left[  \iint\limits_{\check{D}}\rho
rdrd\varphi\right]  \text{ ,} \tag{2.23}%
\end{equation}
we finally obtain,
\begin{equation}
L^{2}(\rho)\ln^{2}\left(  \frac{r_{2}}{r_{1}}\right)  \leq2\pi\ln\left(
\frac{r_{2}}{r_{1}}\right)  \iint\limits_{\check{D}}r\rho drd\varphi\text{
\quad.} \tag{2.24}%
\end{equation}
This produces immediately
\begin{equation}
\frac{L^{2}(\rho)}{A(\rho)}\leq\frac{2\pi}{\ln\left(  \frac{r_{2}}{r_{1}%
}\right)  }\text{ .} \tag{2.25}%
\end{equation}
If we choose $\rho=\frac{a}{2\pi r}$ (where $a$ is some parameter which is
determined below,in Eq.(2.23)),then (2.25) is converted into equality with
$L(\rho)=a$ and $A$($\rho)=\frac{a^{2}}{2\pi}\ln\left(  \frac{r_{2}}{r_{1}%
}\right)  $ .As in the case of a rectangle ,we conclude,that
\begin{equation}
M(\check{D})=\frac{1}{2\pi}\ln\left(  \frac{r_{2}}{r_{1}}\right)  \text{.}
\tag{2.26}%
\end{equation}
Accordingly,by analogy with (2.19) ,we can write
\begin{equation}
A(\check{D})\geq L^{2}(\rho)M(\check{D}) \tag{2.27}%
\end{equation}
which in this case coincides obviously with Eq.(2.25).

Let us observe now that,actually,the rectangle $\hat{R}$ and the annulus
$\check{D}$ could be mapped conformally into each other. Moreover,it can be
shown$[21]$ that $\lambda\left(  \Gamma\right)  $ is conformal invariant.This
can be easily understood if we notice that for any conformal mapping
$z\rightarrow\tilde{z}$ we should have, by construction ,
\begin{equation}
\rho\left|  dz\right|  =\tilde{\rho}\left|  d\tilde{z}\right|  \tag{2.28}%
\end{equation}
and ,accordingly,the area $dA$ is defined by
\begin{equation}
dA=\rho^{2}dx\wedge dy=\rho^{2}\frac{i}{2}dz\wedge d\bar{z}=\tilde{\rho}%
^{2}\frac{i}{2}d\tilde{z}\wedge d\widetilde{z}\text{ .} \tag{2.29}%
\end{equation}
The invariance of $dA$ follows directly from the invariance of the length
element $dl=\rho\left|  dz\right|  $ as can be checked directly from
Eq.(2.29)$[5].$ Clearly,the metric $\rho$ determines all surface properties.In
particular,the singularities of the metric correspond to some surface
singularities (defects) as we shall demonstrate.

In the meantime,let us consider in some detail the conformal mapping of the
rectangle $\hat{R}$ onto $\check{D}$ .To facilitate our understanding ,it is
useful to visualize the mapping,e.g.see Fig.1.\marginpar{Fig.1}The mapping
from $\xi-$plane ($\check{D}$) to z-plane ($\hat{R}$) is being performed by
the following equation:
\begin{equation}
z=a\frac{\ln\xi}{2i\pi}\text{ .} \tag{2.30}%
\end{equation}
It maps the annulus which is cut along the positive $\xi$ axis into the
rectangle .The cut could be avoided if we convert the rectangle $\hat{R}$ into
a cylinder of height $b$ (by identifying [0,$ib]$ with [$a,a+ib]).$The
periodicity is most explicitly seen by rewriting Eq.(2.30) in the form of
\begin{equation}
\xi=\exp\left(  \frac{2i\pi z}{a}\right)  \tag{2.31}%
\end{equation}
so that ,obviously,$\xi(z)=\xi(z+a).$ By combining Eqs.(2.26) and (2.30) we
obtain,
\begin{equation}
M(\check{D})=\frac{1}{2\pi}\ln\left(  \frac{1}{r}\right)  \tag{2.32}%
\end{equation}
where $\frac{1}{2\pi}\ln\left(  \frac{1}{r}\right)  =\frac{b}{a}$ .Using
Eq.(2.27) we obtain,
\begin{equation}
A(\check{D})\geq a^{2}M\text{ ,} \tag{2.33}%
\end{equation}
which coincides \textbf{exactly }with the result (2.19) as anticipated.We
shall need also to use the results for an annulus $\check{D}$ in which some
curve $C$ connects the inner circle $C_{1}$ with the outer circle $C_{2}$
.Simple calculation shows$[21]$
\begin{equation}
\frac{L^{2}(\rho)}{A(\rho)}\leq\frac{1}{2\pi}\ln\left(  \frac{r_{2}}{r_{1}%
}\right)  \text{ .} \tag{2.35}%
\end{equation}
If,as before,we choose $r_{2}=1$ and $r_{1}=r$ ,then we obtain ,
\begin{equation}
\hat{M}=\frac{2\pi}{\ln\left(  \frac{1}{r}\right)  }=\frac{a}{b}\text{ ,}
\tag{2.36}%
\end{equation}
so that ,as before, $A(\check{D})\geq b^{2}\hat{M}=ab.$

The duality between the results (2.23) and (2.36) happens to be very important
as we shall demonstrate in section 4. In the meantime,we need to introduce the
concept of a quadratic differential.This is accomplished in the next section.

\section{Some essentials about quadratic differentials}

Let us begin with Eq.(2.31).Using this equation we obtain the following
sequence of results
\begin{equation}
d\xi=\frac{2i\pi}{a}\xi dz\longrightarrow dz=\frac{a}{2i\pi}\frac{d\xi}{\xi
}\longrightarrow\left(  dz\right)  ^{2}=-\frac{a^{2}}{\left(  2\pi\right)
^{2}}\frac{\left(  d\xi\right)  ^{2}}{\xi^{2}} \tag{3.1}%
\end{equation}
The last expression represents the first example of a quadratic
differential.More formaly,we provide the following

\textbf{Definition} \textbf{3.1}.\textit{ Let }$z=f(\tilde{z})$\textit{ be
some coformal mapping of the domain \~{D} onto D and let }$\varphi(z)$\textit{
be some function wich transforms as}
\begin{equation}
\varphi(\tilde{z})=\varphi(z)\left(  \frac{dz}{d\tilde{z}}\right)  ^{2}\text{
, }z=f(\tilde{z}), \tag{3.2}%
\end{equation}
\textit{then the expression }$\varphi(z)\left(  dz\right)  ^{2}$\textit{ is
called quadratic differential.}

From (3.2) it follows,that this quantity is invariant with respect to mappings
:$z\rightarrow\tilde{z}$ ,i.e.
\begin{equation}
\varphi(z)\left(  dz\right)  ^{2}=\varphi(\tilde{z})\left(  d\tilde{z}\right)
^{2}. \tag{3.3}%
\end{equation}
Eq.(3.1) represents just an example of general result given by
Eq.(3.3).Comparison between Eq.s(2.28) and (3.3) suggests us to introduce

\textbf{Definition} \textbf{3.2}. \textit{The differential }$\left|
dw\right|  =\left|  \varphi(z)\right|  ^{\frac{1}{2}}\left|  dz\right|
$\textit{ is called the lenght element (}$\varphi-$\textit{ length) associated
with }$\varphi.$

\textbf{Remark} \textbf{3.3}.a)\textit{In terms of }$\left|  dw\right|
$\textit{ the length is just the usual Euclidean length ;b)}$\rho(z)$\textit{
can be identified ,in principle, with }$\left|  \varphi(z)\right|  ^{\frac
{1}{2}}$\textit{ but this association is actualy formal as we shall explain
below , in section 4 .}

From the above discussion it follows,in particular,that
\begin{equation}
1\cdot\left(  dw\right)  ^{2}=\varphi(z)\left(  dz\right)  ^{2} \tag{3.4}%
\end{equation}
Suppose,that there is another $\tilde{w}$ such that
\begin{equation}
1\cdot\left(  d\tilde{w}\right)  ^{2}=\varphi(z)\left(  dz\right)  ^{2}.
\tag{3.5}%
\end{equation}
Then,clearly,$\tilde{w}=\pm w+const.$ Moreover,by taking a square root of
Eq.(3.4) we obtain,
\begin{equation}
w=\Phi(z)=\int\limits^{z}dz\sqrt{\varphi(z)}\text{ .} \tag{3.6}%
\end{equation}
From here,we obtain as well
\begin{equation}
\frac{dw}{dz}=\sqrt{\varphi(z)}. \tag{3.7}%
\end{equation}
This is the differential equation on the complex plane or on the Riemann
surface,etc. The flow lines of this equation were used extensively in Part I
of this work.Now ,we would like to provide some additional details. To this
purpose ,let us consider quadratic differential of the type
\begin{equation}
\left(  dw\right)  ^{2}=\left(  \frac{n+2}{2}\right)  ^{2}z^{n}\left(
dz\right)  ^{2} \tag{3.8}%
\end{equation}
where n is some integer (n$\geq-1).$ Use of Eq.(3.6) provides us with the result,%

\begin{equation}
w=\Phi(z)=z^{\frac{n+2}{2}}, \tag{3.9}%
\end{equation}
or
\begin{equation}
\frac{dw}{dz}=\left(  \frac{n+2}{2}\right)  z^{\frac{n}{2}}. \tag{3.10}%
\end{equation}
Eq.(3.9) is a typical example of a conformal mapping.For such mapping ,with
respect to the origin of z-plane, the whole z-plane is subdivided into sectors
(n$\geq-1)$
\begin{equation}
\frac{2\pi}{n+2}k\leq\arg z\leq\frac{2\pi}{n+2}(k+1),k=0,1,...,n+1, \tag{3.11}%
\end{equation}
such that when arg z covers one of these sectors,the resulting image covers
either the upper or the lower w-plane.If we consider the set of parallel
lines(parallel to $\operatorname{Re}w$ axis in $w$-plane), $\operatorname{Im}%
w=an,a=const,n=0,1,2,...,$these parallel lines go into ''parallel lines '' in
one of the sectors of z-plane so that for the Y-type singularity we obtain the
result depicted in Fig.2\marginpar{Fig.2}.In the case of a thorn ,n=-1 ,and we
have the mapping of the whole z-plane (with cut along x$\geq0$ axis) into the
upper w-plane. The horizontals in w-plane go into the ''horizontals'' in
z-plane as depicted in Fig.3.\marginpar{Fig.3}Straightening of the flow lines
explains the reason of the word ''measured foliations'' introduced in Part
I(e.g.see Fig.21 of Part I).From the examples of the previous section it
follows ,that the singularities producing poles of order
$>$%
2 are not acceptable since if we identify $\left|  \varphi(z)\right|  $ with
$\rho^{2}(z),$then the area,Eq.(2.11), becomes divergent.This means,in
particular,that the singularities depicted in Fig.4 and known in the
literature on liquid crystals$[23]$ are mathematicaly
ill-defined\marginpar{Fig.4}(see also the next section for additional
details).The singularities for which n $>1$ are permissible,in principle, and
could be considered along the lines similar to that discussed in Part I (see
Ref.[24] and section 6 below).

All mathematically permissible quadratic differentials can be brought into the
\textbf{standard form} as follows.Consider the mapping
\begin{equation}
\xi(z)=\exp\left(  \frac{2i\pi}{L_{\Phi}}\Phi(z)\right)  . \tag{3.12}%
\end{equation}
It is analogous to Eq.(2.31) and $L_{\Phi}$ is $\varphi-$ length of the
quadratic differential $\Phi(z).$ By analogy with Eq.(3.1) ,we have now
\begin{equation}
d\xi=\frac{2i\pi}{L_{\Phi}}\xi\sqrt{\varphi(z)}dz\leftrightarrows
\varphi(z)\left(  dz\right)  ^{2}=-\frac{L_{\Phi}^{2}}{\left(  2\pi\right)
^{2}}\frac{\left(  d\xi\right)  ^{2}}{\xi^{2}} \tag{3.13}%
\end{equation}
The flow lines of such defined quadratic differential are the
\textbf{concentric circles} (Fig.4 ,Part I).This is very important result
since it allows to map an arbitrary mathematicaly permissible quadratic
differential into the standard differential,Eq.(3.1) (or (3.13)),so that the
rest of the arguments related to the annulus and the rectangle presented in
section 2 could be carried through without change.

The above results can be broadly generalized now.To this purpose let us take
another look at Eq.(3.12).What we actually have is a mapping from some Riemann
surface R on which the quadratic differential ''lives' into the surface of the
flat annulus.Clearly,such mapping has some limitations.That is , it might as
well be that the above mapping exist only for some ring domain $\left|
z\right|  =\rho$ on R .The size of this domain determines the size of the
annulus (or punctured disk).Without going into intricate details about the
correspondence of thee domains$[25,26]$we provide

\textbf{Theorem 3.4}.\textit{If }$\varphi$\textit{ is quadratic differential
which can have closed trajectories on R with respective lengths }%
$L_{\varphi_{i}}$\textit{ ,then its characteristic ring domains (i.e.the
maximal ring domains swept out by the closed trajectories cover R up to a set
of measure zero.For a holomorphic quadratic differential on a compact surface
of genus g}$\geq2$\textit{ the number of characteristic ring domains is at
most 3g-3.}

\textbf{Proof}.Please,consult Refs.[7,8,25,26]. $\square$

Let us explain what all this actually means.It is well known $[27],$ that
every Riemann surface R could be decomposed into set of 2g-2
pants.Conversely,if we have at our disposal a set of 2g-2 pants ,we can
restore R if we have a 3-valent planar graph with 2g-2 vertices and 3g-3 edges
which does not have free ends.3g-3 edges correspond to cylinders made when the
pants are glued together.All this is depicted in Fig.5.\marginpar{Fig.5}Since
we have discussed already the problems which involve cylinders,e.g.see Fig.1
,it becomes clear why we have discussed them in the first place.Moreover,the
ring domains form the union of closed geodesics(lamination ,according to the
Definition 4.1.of Part I),one for each of these cylinders$[27-29]$ .And,when
these geodesics are projected into the open disk D$^{2},$as discussed in
section 4 ,Part I, we obtain the projective meanders.

With this background ,we are ready now for some applications of these results..

\section{Applications of quadratic differentials to textures in liquid crystals}

In the previous sections we have provided essential mathematical background
related to quadratic differentials.Before we actualy use them (in this and the
following sections) it is desirable to provide some relevant physical
background related to quadratic differentials.

\subsection{Differential geometric meaning of quadratic differentials}

We have introduced the length ,Eq.(2.28),and the area ,Eq.(2.29), and
identified $\left|  \varphi(z)\right|  ^{\frac{1}{2}}$ \quad with $\rho$ and
$\left|  \varphi(z)\right|  $ with $\rho^{2}$ .This,however,is not enough.Here
we shall explain why.

Although the Cartan method of description of surfaces is the most elegant and
economical,moreover,it is indispensable for the discussion of general
topological properties of surfaces$[30]$ ,we shall avoid its use here for the
reasons which will become obvious upon reading.

Let \textbf{r }be the point in $R^{3}$ which belongs to some surface $S.$
Then,on one hand, \textbf{r} is just some 3d Euclidean vector with components
r$_{1}$ ,r$_{2}$ ,and r$_{3}$ and,on the other hand,because it belongs to the
surface ,its location on the surface could be given in terms of some local
coordinates x$_{1},$x$_{2}\in S.$ With such defined \textbf{r }it is useful to
construct two vectors $\frac{d\mathbf{r}}{dx_{1}}$ and $\frac{d\mathbf{r}%
}{dx_{2}}$ so that the induced metric of the surface is given by
\begin{equation}
dl^{2}=Edx_{1}^{2}+2Fdx_{1}dx_{2}+Gdx_{2}^{2} \tag{4.1}%
\end{equation}
where $E=\left(  \frac{d\mathbf{r}}{dx_{1}}\right)  ^{2},F=\frac{d\mathbf{r}%
}{dx_{1}}\cdot\frac{d\mathbf{r}}{dx_{2}},G=\left(  \frac{d\mathbf{r}}{dx_{2}%
}\right)  ^{2}.$ Introduce now the unit normal vector to the surface
$\mathbf{T}$ via
\begin{equation}
\mathbf{T=}\frac{\frac{d\mathbf{r}}{dx_{1}}\wedge\frac{d\mathbf{r}}{dx_{2}}%
}{\left|  \frac{d\mathbf{r}}{dx_{1}}\wedge\frac{d\mathbf{r}}{dx_{2}}\right|
}\text{ \quad,} \tag{4.2}%
\end{equation}
then the second fundamental form of surface can be written as
\begin{equation}
\text{II}=Ldx_{1}^{2}+2Mdx_{1}dx_{2}+Ndx_{2}^{2} \tag{4.3}%
\end{equation}
where $L=-\frac{d\mathbf{T}}{dx_{1}}\cdot\frac{d\mathbf{T}}{dx_{1}}%
,M=-\frac{d\mathbf{T}}{dx_{1}}\cdot\frac{d\mathbf{r}}{dx_{2}}$ and
$N=-\frac{d\mathbf{T}}{dx_{2}}\cdot\frac{d\mathbf{r}}{dx_{2}}$ \quad.

In terms of the quantities just defined one can determine the Gauss(intrinsic
) curvature K as
\begin{equation}
K=k_{1}k_{2}=\frac{LN-M^{2}}{\hat{\lambda}^{2}} \tag{4.4}%
\end{equation}
and the mean (extrinsic) curvature $H$ as
\begin{equation}
H=\frac{1}{2}\left(  k_{1}+k_{2}\right)  =\frac{L+N}{2\hat{\lambda}} \tag{4.5}%
\end{equation}
where the factor $\hat{\lambda}$ is associated with the first fundamental form
which is brought into the conformal form
\begin{equation}
dl^{2}=\hat{\lambda}(dx_{1}^{2}+dx_{2}^{2}) \tag{4.6}%
\end{equation}
and $k_{1}$ and $k_{2}$ are the principal curvatures of the surface.

The displacement of the vector \textbf{r }along some curve which belongs to S
is determined by the vector \textbf{t },i.e.
\begin{equation}
\mathbf{t=}\frac{d\mathbf{r}}{dl}=\frac{dx_{1}^{{}}}{dl}\cdot\frac
{d\mathbf{r}}{dx_{1}}+\frac{dx_{2}}{dl}\cdot\frac{d\mathbf{r}}{dx_{2}}\text{
.} \tag{4.7}%
\end{equation}
According to the rules of differential geometry of curves$[12]$ ,we have
\begin{equation}
\frac{d\mathbf{t}}{dl}=\kappa\mathbf{T} \tag{4.8}%
\end{equation}
where $\kappa$ is the local curvature of the \textbf{curve }which belongs to
S.Depending upon how the line is drawn on the surface , $\kappa$ may vary from
$k_{1}$ to $k_{2}$ (if $k_{1}>k_{2}).$ The lines along which $\kappa=k_{1}($or
$k_{2})$ are called the principal curvature directions.They form an orthogonal
network.The equation for the lines of principal curvatures is given by
\begin{equation}
-Mdx_{1}^{2}+(L-N)dx_{1}dx_{2}+Mdx_{2}^{2}=0\text{ .} \tag{4.9}%
\end{equation}
Finally,the Codazzi equations (the consistency equations) could be written as
follows
\begin{equation}
\frac{d}{dx_{1}}\left(  \frac{L-N}{2}\right)  +\frac{d}{dx_{2}}M=\hat{\lambda
}\frac{d}{dx_{1}}H \tag{4.10a}%
\end{equation}%
\begin{equation}
\frac{d}{dx_{2}}\left(  \frac{L-N}{2}\right)  -\frac{d}{dx_{1}}M=-\hat
{\lambda}\frac{d}{dx_{2}}H\text{ .} \tag{4.10b}%
\end{equation}
Introduce now the complex notations: $z=x_{1}+ix_{2}$ and take into account
that for an arbitrary complex function $F(z,\bar{z})=F_{1}+iF_{2}$ one can
write
\[
2\frac{d}{dz}F=\left(  \frac{d}{dx_{1}}F_{1}+\frac{d}{dx_{2}}F_{2}\right)
-i\left(  \frac{d}{dx_{2}}F_{1}-\frac{d}{dx_{1}}F_{2}\right)
\]
and another equation which is the complex conjugate of this.Then,following
Hopf$[4]$ ,we can introduce the Hopf differential
\begin{equation}
\Phi(z,\bar{z})=\frac{L-N}{2}-iM\text{ .} \tag{4.11}%
\end{equation}
It could be shown ,that
\begin{equation}
\frac{\left|  \Phi\right|  }{\lambda}=\left|  \frac{k_{1}-k_{2}}{2}\right|  .
\tag{4.12}%
\end{equation}
That is \textbf{the umbilic points }($k_{1}=k_{2})$ \textbf{are zeros of
}$\Phi$ .With help of $\Phi$ both of the Codazzi equations could be rewritten
in the compact form given by
\begin{equation}
\frac{\partial\Phi}{\partial\bar{z}}=\hat{\lambda}\frac{\partial}{\partial
z}H\text{ .} \tag{4.13}%
\end{equation}
Moreover,$\Phi$ itself can be also rewritten in more compact form as
\begin{equation}
\Phi=-2\frac{d\mathbf{r}}{dz}\cdot\frac{d\mathbf{T}}{dz}\text{ .} \tag{4.14}%
\end{equation}
Let us consider now the change of parametrization of the surface,i.e.
$x_{1},x_{2}\longrightarrow x_{1}^{\prime},x_{2}^{\prime}$ and let $\xi
=x_{1}^{\prime}+ix_{2}^{\prime}.$ Clearly,we expect
\begin{equation}
\frac{d\mathbf{r}}{dz}=\frac{\partial\mathbf{r}}{\partial\xi}\frac{d\xi}{dz}
\tag{4.15a}%
\end{equation}
and
\begin{equation}
\frac{d\mathbf{T}}{dz}=\frac{\partial\mathbf{T}}{\partial\xi}\frac{d\xi}%
{dz}\text{ .} \tag{4.15b}%
\end{equation}
The combined use of Eq.s (4.14) and (4.15) produces
\begin{equation}
\Phi(z,\bar{z})\left(  dz\right)  ^{2}=\Phi(\xi,\bar{\xi})\left(  d\xi\right)
^{2} \tag{4.16}%
\end{equation}
which coincides with the transformation rule for the quadratic
differentials,Eq.(3.3). \ \ \ \ \ \ Hence $\Phi$ \textbf{is} the quadratic
differential! Moreover,the equation for the principal curvatures, Eq.(4.9)
,can now be rewritten in compact suggestive form as
\begin{equation}
\operatorname{Im}\left[  \Phi\left(  dz\right)  ^{2}\right]  =0 \tag{4.17}%
\end{equation}
which is equivalent to
\begin{equation}
\arg\Phi+2\arg\left(  dz\right)  =m\pi,\text{ \quad}m=0,\pm1,...\text{ \quad.}
\tag{4.18}%
\end{equation}
This result can be conveniently rewritten as
\begin{equation}
\arg\left(  dz\right)  =\frac{m\pi}{2}-\frac{1}{2}\arg\Phi\text{ .} \tag{4.19}%
\end{equation}
Let we have now the isolated umbilic ($k_{1}=k_{2})$ point $p$ on the
surface.Let us surround this point by some closed nonselfintersecting contour
$\mathcal{C}$ and let us define the index of such point $I(p)$ as
\begin{equation}
I(p)=\frac{1}{2\pi}\delta\left(  \arg\left(  dz\right)  \right)  \tag{4.20}%
\end{equation}
where $\delta$ means the variation of an angle when one is circling around
$\mathcal{C}$ counterclockwise.Combining of Eq.s(4.19) and (4.20) produces
\begin{equation}
I(p)=-\frac{1}{2\pi}\frac{1}{2}\delta\left(  \arg\Phi\right)  \text{ .}
\tag{4.21}%
\end{equation}
Assume now that locally $\Phi$ can be written as
\begin{equation}
\Phi(z,\bar{z})=cz^{n}+...\text{ \quad.} \tag{4.22}%
\end{equation}
For the umbilic point $n>0$ and $\hat{\lambda}$ is nonsingular for $k_{1}\neq
k_{2}.$ In addition,for $k_{1}\neq k_{2}$ we may have singular $\hat{\lambda}$
(then we may have $n<0).$ To calculate the index of such singularity
explicitly,we use known fact of complex analysis $[31]$%
\begin{equation}
\delta(\arg\Phi)=\operatorname{Im}\oint\limits_{\mathcal{C}}\frac{d\Phi}{\Phi
}\text{ .} \tag{4.23}%
\end{equation}
This produces,in view of Eq.s (4.21) and (4.23),
\begin{equation}
I(p)=-\frac{n}{2}\text{ , \quad}n=\pm1,\pm2,...\text{ \quad.} \tag{4.23a}%
\end{equation}

\textbf{Definition} \textbf{4.1}.\textit{We shall call Eq.(4.23a) the Hopf
quantization rule }.

This result for the index produces exactly the results obtained for the liquid
crystals,e.g. see section 1(Part I) or Ref.23. In the case of Y-type
singularity ,we have, according to Strebel$[7],n=+1,$which produces
$I_{Y}=-\frac{1}{2}$ in accord with Fig.2 of Part I. For the case of a
thorn,we have $n=-1,$ which produces $I_{I}=\frac{1}{2}$ again, in complete
accord with Fig.2 of Part I and with Refs.[12,32].The index for the rest of
singularities can be computed now analogously.

\textbf{Remark} \textbf{4.2}. \textit{Poincare}$[39]$\textit{ had originally
considered singularities of differential equations of the form}
\begin{equation}
a(x,y)dx+b(x,y)dy=0 \tag{4.24}%
\end{equation}
\textit{This is just the condition of orthogonality between the vector
(}$a,b)$\textit{ and the curve (}$dx,dy).$\textit{ By going around a closed
curve }$C$\textit{ the direction of the vector (}$a,b)$\textit{ can change
only by 2}$\pi n$\textit{ where }$n$\textit{ is an integer.Hence,by having
half integer values for }$I$\textit{ is equivalent of not having vector fields
on the surfaces.This was actually stated in Part I(section 1) without proof.}

\subsection{Quadratic differentials and the textures in liquid crystals}

The fact that the textures in liquid crystals can be associated with measured
foliations was recognized already by Poenaru$[3]$and Langevin $[4]$ .The fact
that the measured foliations are asssociated with quadratic differentials was
explained by Hubbard and Masur$[2].$A comprehensive treatment of measured
foliations could be found in Refs.24,35.Here ,we only provide some details to
make this identification complete.Before doing so ,we would like to mention
that the results below are equally applicable to any kind of ''hyperbolic
paper''(the terminology used by Thurston $[36]$ )that is to any surface which
may develop some crumples(see also section 7).

The free energy of distortion $\mathcal{F}_{d}$ in the case of nematics (in
one constant approximation) is given by $[37]$
\begin{equation}
\mathcal{F}_{d}=\frac{1}{2}\tilde{K}\int d^{3}r\left(  \nabla\mathbf{n}%
\right)  ^{2} \tag{4.24}%
\end{equation}
where the coupling constant $\tilde{K}$ has dimensionality energy/cm and is
related effectively to the surface tension.The coupling constant $\tilde{K}$
exibits therefore noticeable temperature dependence.The unit vector
\textbf{n}=\textbf{n}(\textbf{r}) is known in the literature on liquid
crystals as \textbf{director}. On one hand, it is being used to describe the
orientation of the individual molecule (usually very stiff, rod-like, organic
molecule of not too high molecular weight)with respect to some preassigned
fixed axis,on another hand, \textbf{n }in Eq.(4.24) is also called a director
although it is no longer is directly associated with the individual
molecule.It is possible, however,to obtain the macroscopic distortion energy
$\mathcal{F}_{d}$ from the underlying microscopic molecular model of nematic
liquid crystal$[38]$ .One is usually looking for a minimum of $\mathcal{F}%
_{d}^{{}}$ under the additional constraint :\textbf{n}$^{2}=1($the nonlinear
sigma model$[39]).$ In mathematical literature $[40,41]$ ,the same problem is
stated somewhat more precisely.. Specifically,let $\varphi(\mathbf{x)=}%
\frac{\mathbf{x}}{\left|  \mathbf{x}\right|  }=\mathbf{n,x\in}%
\mathcal{U\subset}\mathbf{R}^{3}$ (the domain $\mathcal{U}$ may coincide with
\textbf{R}$^{3}).$Let \{$H_{i}\}$ be k disjoint compact subsets of
$\mathcal{U}$ which are called ''holes''.$\varphi(\mathbf{x})$ is the Gauss
map $\varphi:\mathcal{U\rightarrow}\mathit{S}^{2}$ in the absence of holes.In
the presence of holes,consider a spherical neighborhood around some particular
$H_{i}.$ If $\varphi$ is restricted to this neighborhood,then,with such
restriction,$\varphi$ defines a map $\mathit{S}^{2}\longrightarrow S^{2}.$
This map has a degree $d_{i}\in\mathbf{Z}$ (integers,possibly including
zero).If now $\Omega=\mathcal{U}$%
$\backslash$%
($\bigcup\limits_{i=1}^{k}H_{i}),$ then one is interested in finding the
harmonic map $\varphi:\Omega\rightarrow S^{2}$ ,that is to find
\begin{equation}
E=\inf_{\varphi\in\varepsilon}\int\limits_{\Omega}d^{3}x\left(
\bigtriangledown\varphi\right)  ^{2} \tag{4.25}%
\end{equation}
under conditions (i=1-k)
\[
\varepsilon=\{\varphi\in(\Omega;S^{2})\mid\deg(\varphi;H_{i})=d_{i},\text{
}\int\limits_{\Omega}d^{3}x\left(  \bigtriangledown\varphi\right)  ^{2}%
<\infty\}.
\]
The above problem can be formulated as well in two dimensions.In this case
,one is talking about the harmonic maps from $S^{1}$ to $S^{1}$.For $d_{i}$
which are \textbf{integers} and satisfy the condition $\sum\limits_{i}d_{i}=0$
the problem is solved in Ref[42] with the result :
\begin{equation}
E=\frac{1}{2}\int\limits_{\Omega}d^{2}x\left(  \bigtriangledown\varphi\right)
^{2}=-\pi\sum\limits_{i<j}d_{i}d_{j}\ln\left|  a_{i}-a_{j}\right|
+\text{boundary term +const} \tag{4.26}%
\end{equation}
where $\left\{  a_{i}\right\}  $ play the same role as $\left\{
H_{i}\right\}  $ in 3 dimensions.This result provides the desired Coulomb gas
analogy used for the description of the defect mediated melting transitions in
quasi two dimensional liquid crystals $[43]$ via the Kosterlitz-Thouless type
of transition which was discussed in Part I.%

$>$%
From \textbf{the same }physical literature it is known,however,$[23,37,43]$
that the defects with integer $d_{i}$ are topologicaly unstable (at least in
\textbf{R}$^{3})$ and only those which have the half integer $d_{i}$ are
topologicaly stable.

\textbf{Remark} \textbf{4.3}.\textit{ In two dimensions the degree }$d_{i}%
$\textit{ coincides with the index} $I.$

\textbf{Remark} \textbf{4.4}.\textit{In two dimensions ,in view of the Remark
4.1.,if we would have stable defects with integer and halfinteger }$I$\textit{
we would run into problem since the textures originating from defects with
integer }$I$\textit{ produce the orientable (vector) fields on the surface
while the defects with half integer }$I$\textit{ produce the non-orientable
(line) fields.According to Strebel's book,Ref.7,in this case one should
consider the field of textures (foliations) coming from defects with integer
index }$I$\textit{ as non-orientable line field.\footnote{*In the case of 2+1
gravity ,section 5,such remark is equivalent of saying that use of quadratic
differentials allows to provide a formal unification of the description of
electicity and gravity .Such unification is completely different in nature
from that proposed by Kaluza and Klein.Please,see also section 6.2.}*}

The quadratic differential,Eq.(3.1) ,describes the defect with $I=1$ and is
depicted in Fig.4b) of Part I,while the same differential with $a\rightarrow
ia$ is also having index $I=1$ and is depicted in Fig.2 of Part I.The very
existence of the Whitehead moves , Fig.4 ,Part I,would be questionable should
\textbf{both} integer and halfinteger fields not be treated as the line fields$[7].$

As in the case of 2+1 gravity (to be discussed in section 5),because the line
fields are nonorientable,there is no interaction between the defects but there
is an energy,nevertheless. And this energy can be minimized.

Indeed,in two dimensions we have \textbf{n=}$\mathbf{\{}\cos\varphi
(\mathbf{r}),\sin\varphi(\mathbf{r})\},\mathbf{r=}\{x,y\}.$ This
produces,instead of Eq.(4.24),the following result for the distortion energy
\begin{equation}
\mathcal{F}_{d}=\frac{\tilde{K}}{2}\int\limits_{\Omega}d^{2}r\left(
\bigtriangledown\varphi\right)  ^{2} \tag{4.27}%
\end{equation}
where the domain $\Omega$ is analogous to that defined in Eq.(4.25) (adjusted
for 2-dimensional case).The functional $D[\varphi]=\frac{2}{K}\mathcal{F}_{d}$
is known in the literature as Dirichlet integral$[44].$ This integral has some
remarkable properties summarized in the following theorems

\textbf{Theorem} \textbf{4.5}.\textit{Let the function }$w=f(z)$\textit{
provide the conformal mapping of the domain }$\Omega$\textit{ onto }%
$\Omega^{\ast}$\textit{ and let }$\varphi(x,y)=\psi(f(z)),$\textit{ then}
\begin{equation}
D[\varphi]=\iint\limits_{\Omega}\left(  \varphi_{x}^{2}+\varphi_{y}%
^{2}\right)  dxdy=\iint\limits_{\Omega^{\ast}}\left(  \psi_{u}^{2}+\psi
_{v}^{2}\right)  dudv \tag{4.28}%
\end{equation}
\textit{where }$z=x+iy,w=u+iv$\textit{ and }$\varphi_{x}=\frac{\partial
\varphi}{\partial x}$\textit{ ,etc.}

\textbf{Proof}.Indeed,taking into account that
\begin{equation}
\varphi_{x}=\psi_{u}u_{x}+\psi_{v}v_{x} \tag{4.29a}%
\end{equation}%
\begin{equation}
\varphi_{y}=\psi_{u}u_{y}+\psi_{v}v_{y} \tag{4.29b}%
\end{equation}
and empoloying the Cauchy-Riemann equations
\begin{equation}
u_{x}=v_{y}\text{ ,}v_{x}=-u_{y} \tag{4.30}%
\end{equation}
we obtain,
\begin{equation}
\varphi_{x}^{2}+\varphi_{y}^{2}=\left(  \psi_{u}^{2}+\psi_{v}^{2}\right)
\cdot\left(  u_{x}^{2}+u_{y}^{2}\right)  \text{ .} \tag{4.31}%
\end{equation}
But the Jacobian $J=u_{x}v_{y}-u_{y}v_{x}=u_{x}^{2}+u_{y}^{2}$ in view of
Eq.(4.30).Hence, indeed, $D[\varphi]$ is conformal invariant.$\square$

\textbf{Corollary} \textbf{4.6}.\textit{Taking into account that}
\begin{equation}
\left(  \varphi_{x}^{2}+\varphi_{y}^{2}\right)  dxdy=4\left|  \varphi
_{z}(z)\right|  ^{2}d^{2}z \tag{4.32}%
\end{equation}
\textit{one has actualy more:}
\begin{equation}
\iint\limits_{\Omega}d^{2}z\left|  \varphi_{z}(z)\right|  ^{2}=\iint
\limits_{\Omega^{\ast}}d^{2}w\text{ .} \tag{4.33}%
\end{equation}
\textit{That is the Dirichlet integral }$D[\varphi]$\textit{ defined in
}$\Omega$\textit{ is equal to the area of }$\Omega^{\ast}$\textit{ which is
the image of the area }$\Omega$\textit{ upon the conformal mapping }$w=f(z)$\textit{.}

\textbf{Theorem} \textbf{4.7}.\textit{Let }$w=\varphi(z)$\textit{ is the
conformal mapping and }$\Phi(w)$\textit{ is the quasiconformal mapping ,that
is it is performed with help of function }$\Phi$\textit{ such that}
\begin{equation}
d\Phi=\Phi_{w}dw+\Phi_{\bar{w}}d\bar{w}\text{ ,} \tag{4.34}%
\end{equation}
\textit{then}
\begin{equation}
D[\Phi(\varphi(z))]\leq\hat{K}D[\varphi(z)]\text{ ,} \tag{4.35a}%
\end{equation}
\textit{where the dilatation factor }$\hat{K}$\textit{ is defined by}
\begin{equation}
\hat{K}=\frac{\left|  \Phi_{w}\right|  +\left|  \Phi_{\bar{w}}\right|
}{\left|  \Phi_{w}\right|  -\left|  \Phi_{\bar{w}}\right|  }\text{ , }\hat
{K}>1\text{ \quad if \quad}\left|  \Phi_{\bar{w}}\right|  \neq0. \tag{4.35b}%
\end{equation}

\textbf{Proof}.Please,consult Ref.21.$\square$

\mathstrut

In view of Theorem 4.5 ,our use of isoperimetric inequalities ,discussed in
section 2,and quadratic differentials ,discussed in section 3 ,becomes almost
obvious.Since $D[\varphi]$ has the meaning of an area , we can write as well
(in view of Eq.(2.29)):
\begin{equation}
D[\varphi]=\iint\limits_{\Omega}\rho^{2}(\varphi)dx\wedge dy \tag{4.36}%
\end{equation}
where now
\begin{equation}
\rho^{2}(\varphi)=\varphi_{x}^{2}+\varphi_{y}^{2}\text{ .} \tag{4.37}%
\end{equation}
As for the length $L_{\varphi}$ ,it should be defined accordingly through
$\left|  dw\right|  $=$\rho\left|  dz\right|  .$ Then,the isoperimetric
inequalities discussed in section 2 can be used immediatly.As for the Theorem
4.7.,we shall need it later,in section 6,when we shall discuss some more
advanced topics.For the time being ,we need to discuss how to obtain $\rho$
from $\rho^{2}$ given by Eq.(4.37).To this purpose ,following Ref[8] ,let us
introduce the function $g_{z}=\frac{1}{2}\left(  g_{x}-ig_{y}\right)  $ and
such that $g=u+i\varphi$ .Furthermore,let
\begin{equation}
g_{z}=\sqrt{\Phi(z)} \tag{4.38}%
\end{equation}
where the above equation defines $\Phi.$ Consider now
\begin{equation}
\iint\limits_{\Omega}\left|  \Phi\right|  dxdy=\iint\limits_{\Omega}\left|
g_{z}\right|  ^{2}dxdy\text{ .} \tag{4.39}%
\end{equation}
Consider as well the contour integral along some closed curve $\gamma\in
\Omega$ given by
\begin{equation}
\oint\limits_{\gamma}\left|  \operatorname{Im}(\sqrt{\Phi}dz)\right|
=\frac{1}{2}\oint\limits_{\gamma}\left|  \left(  \varphi_{x}-u_{y}\right)
dx+\left(  u_{x}-\varphi_{y}\right)  dy\right|  \text{ \quad.} \tag{4.40}%
\end{equation}
If the function $g_{z}$ is analytic(holomorphic),then the Cauchy-Riemann
equations produce :$u_{x}=\varphi_{y}$ and $u_{y}=-\varphi_{x}$ which leads us
to the conclusion that
\begin{equation}
\oint\limits_{\gamma}\left|  \operatorname{Im}(\sqrt{\Phi}dz)\right|
=\oint\limits_{\gamma}\left|  d\varphi\right|  \tag{4.41}%
\end{equation}
where $\left|  d\varphi\right|  =\left|  \varphi_{x}dx+\varphi_{y}dy\right|
.$ At the same time,the same Cauchy-Riemann equations applied to Eq.(4.39)
produce
\begin{equation}
D[\varphi]=\iint\limits_{\Omega}\left(  \varphi_{x}^{2}+\varphi_{y}%
^{2}\right)  dxdy=A_{\Omega}(\rho).\text{ } \tag{4.42}%
\end{equation}
Although thus defined construction had provided us with the area $A_{\Omega
}(\rho)$ ,the contour integral
\begin{equation}
h_{\varphi}(\gamma)=\oint\limits_{\gamma}\left|  d\varphi\right|  \tag{4.43}%
\end{equation}
is \textbf{not exactly }the $\varphi-$length.It is called the \textbf{height}%
(the $\varphi-$length is just $\oint\limits_{\gamma}\left|  \sqrt{\Phi
}\right|  \left|  dz\right|  $).If we are looking for the extremum of (4.42)
,we cannot apply directly the results of section 2.To find out the geometric
meaning of the height ,it is sufficient to go back to Fig.1,and to use the
Remark 3.1. along with Eq.(3.4).Since $\left(  dw\right)  ^{2}$ in Eq.(3.4) is
just the usual Euclidean length element ,this means ,that in terms of $w$ we
have to consider some square ,e.g. like that depicted in Fig.1,with
$\operatorname{Im}w=b$ being indeed the height of the square.Eq.(4.41)
reflects just this fact.Suppose ,as in Fig.1 ,we have an annulus and inside of
the annulus we have a closed curve which touches both the inner and the outer
circle.Then,the image of this curve in w-plane will also be a closed curve
which touches the horizontals .This curve,naturaly,will have some horizontal
and vertical parts.If the height of the rectangle is $\ b$,then we obtain ,
\begin{equation}
h_{\varphi}(\gamma)=b\leq\int\limits_{b_{u}}\left|  d\tilde{\varphi}\right|
=\int\limits_{b}^{a}\left|  \frac{\partial\tilde{\varphi}(u,\varphi)}%
{\partial\varphi}\right|  d\varphi\tag{4.44}%
\end{equation}
where $\tilde{\varphi}(u,\varphi)$ denotes any smooth curve which joins the
horizontals and $b_{u}$ represents the contour.Multiplying both sides of
(4.44) by $a$ we obtain,
\begin{equation}
ab\leq\int\limits_{0}^{a}\int\limits_{0}^{b}\left|  \frac{\partial
\tilde{\varphi}(u,\varphi)}{\partial\varphi}\right|  d\varphi du\text{ .}
\tag{4.45}%
\end{equation}
By squaring the above inequality and using the Schwarz inequality (analogous
to (2.15))we obtain,
\begin{equation}
\left(  ab\right)  ^{2}\leq ab\int\limits_{0}^{a}\int\limits_{0}^{b}\left(
\frac{\partial\tilde{\varphi}}{\partial\varphi}\right)  ^{2}d\varphi du\leq
ab\int\limits_{0}^{a}\int\limits_{0}^{b}\left[  \left(  \frac{\partial
\tilde{\varphi}}{\partial\varphi}\right)  ^{2}+\left(  \frac{\partial
\tilde{\varphi}}{\partial u}\right)  ^{2}\right]  d\varphi du\text{ .}
\tag{4.46}%
\end{equation}
Finaly,using the Theorem 4.5. ,we obtain ,
\begin{equation}
ab\leq\iint\limits_{\Omega}\left[  \tilde{\varphi}_{x}^{2}+\tilde{\varphi}%
_{y}^{2}\right]  dxdy\text{ .} \tag{4.47}%
\end{equation}
Of course,in the above derivation we had made a restriction of having just one
puncture in $\Omega.$We have effectively surrounded the puncture (the critical
point or the singularity) by a circle and considered the domain $\Omega$ as an
annulus which was converted into the rectangle (as usual).Evidently,the
generalization to many singularities should be obvious now.The combination
$ab$ can be written therefore as follows:
\begin{equation}
ab=b^{2}\hat{M}=b^{2}/M=a^{2}M=a^{2}/\hat{M} \tag{4.48}%
\end{equation}
,where $M$ is given by Eq.(2.26) and $\hat{M}=M^{-1}$ according to
eq.(2.35).Using these results \ we can write our final expression for the
distortion energy, Eq.(4.27), for a set of punctures:
\begin{equation}
\mathcal{F}_{d}\geq\frac{\tilde{K}}{2}\sum\limits_{i}a_{i}^{2}M_{i} \tag{4.49}%
\end{equation}
where $a_{i}$ is equal to $\left|  I(p_{i})\right|  $ as can be seen from
Eq.(4.23)(or page 14(top)of Ref.[21]). Additional details related to the
choice of domains,structure of trajectories of quadratic
differentials,etc.,could be found in Ref.[45] which does not contain any
physical applications however .Eq.(4.49) coincides with Eq.(5.3) (see also
Eq.(2.5))of Part I where it was given without proof.

\textbf{Remark} \textbf{4.8}.\textit{The existence of the hexagonally ordered
phases discussed in Part I,e.g. see Figs 8 and 10 of Part I,can be easily
understood based on the results just obtained.Indeed,let us surround each
defect with simple (non self-intersecting) contour }$C$\textit{ and consider
an area which is enclosed by such contour.Now ,the problem can be formulated
as follows:for a given perimeter length }$l$\textit{ of the contour }%
$C$\textit{ find a minimal area }$A$\textit{ which such contour encloses
,provided that the obtained figure can cover the surface }$S$\textit{ without
gaps (i.e. tesselates }$S).$\textit{ For R}$^{2}$\textit{ , }$S^{2}%
$\textit{and R the results are well known}$[46].$\textit{In particular,for
R}$^{2}$\textit{ there are only two options : to have squares or to have
equilateral triangles.For given perimeter length }$l$\textit{ the area }%
$A$\textit{ of the triangle is smaller than that of the square.Hence,
triangles tesselate R}$^{2}$\textit{ under the most optimal conditions and
this is the origin of the existence of the hexatic phase .For the alternative
(physical) proof of the existence of the hexatic phase, please,consult Ref.[47].}

\section{Applications to 2+1 gravity}

Although in the previous sections we have provided all necessary essentials
needed for the description of classical 2+1 gravity,here we need some
ramifications of the obtained results to facilitate the uninterrupted reading.

\subsection{Conical singularities and quadratic differentials}

The connection between the conical singularities and quadratic differentials
was discovered by Troyanov$[10].$His derivation is incomplete,however,as was
recently noticed by Rivin$[11].$ Because of this incompleteness,we would like
to provide here an alternative derivation of Troyanov's results in order to
make connections with that of Rivin.

The main idea of both works lies in the recognition of the fact that any 2
dimensional Riemann surface (and also 3 dimensional manifold $[36])$ admits
consistent triangulation with help of \textbf{flat Euclidean} triangles.The
curvature effects are concentrated then on vertices (cones) of the
triangulated surface. If the surface is without boundary(closed),then the
curvature singularities are well modelled by the cones on flat Euclidean
backgroundsIf the surface has boundaries,then some of the singularities should
be modelled by the punctured(truncated) cones$[11,48]$ .Surfaces which have
negative Euler characteristics are hyperbolic and,whence,they represent an
example of a ''hyperbolic paper''(in Thurston's terminology$[36])$ .Since the
hyperbolicity is always associated with tractrix-like conical -type
surfaces(e.g.see Fig.6)$[17,36]$ and since such surfaces \textbf{cannot }be
smoothly embedded into 3 dimensional Euclidean space,e.g.see Theorem 7.1.
below and Refs[ 36,46],the very tip of the cone may be cut off. This creates a
boundary ( for an illustration,please,consult Ref.[48],especialy pages
98,99)since the cone is being truncated now.The Euler characteristic of such
triangulated surfaces with truncated cones could be calculated with some
effort$[11]$ thus providing the major correction to the results of
Troyanov.This correction ,naturaly,is affecting the results related to 2+1
gravity as we shall demonstrate shortly below.Consider now the following

\textbf{Lemma} \textbf{5.1}.\textit{Suppose we have the quadratic differential
which has the infinitesimal length }$dl^{2}$\textit{ given by }%
\begin{equation}
dl^{2}=\left|  z\right|  ^{2\beta}\left|  dz\right|  ^{2}\text{ , }\beta
\geq-1, \tag{5.1}%
\end{equation}
\textit{then , there is a conical metric given by }%
\begin{equation}
dl^{2}=dr^{2}+r^{2}dt^{2}\text{ ,0}\leq t\leq\alpha2\pi\text{ ,0}\leq
\alpha\leq1, \tag{5.2}%
\end{equation}
\textit{so that (5.1) can be mapped into (5.2) provided that }$\alpha=1+\beta
$\textit{ and }$\alpha=\frac{\theta}{2\pi}$\textit{ with }$\theta$\textit{
being an angle of the cone.}

\textbf{Proof}.We had seen already in section 3,e.g.see Eq.s (3.9),(3.11)and
Fig.2. ,that the maping of the type w=z$^{\frac{n+2}{2}}$ converts the sector
0$\leq\arg z\leq\frac{2\pi}{n+2}$ in z-plane into the upper w half plane.If we
identify the sides of this sector ,we shall obtain a cone with an angle
$\frac{2\pi}{n+2}.$ Clearly,the metric $dl^{2\text{ }}$in w-plane is
Euclidean,i.e.$dl^{2}$=$dx^{2}+dy^{2}=dr^{2}+r^{2}d\varphi^{2}.$ When going
from w to z-plane ,we anticipate that the metric will have the form given by
Eq.(4.6).To calculate the conformal factor $\hat{\lambda}$ ,let us consider in
the light of just presented example the transformation:%

\begin{equation}
x=ar^{\alpha}\cos\alpha\varphi\text{ , }y=ar^{\alpha}\sin\alpha\varphi\text{
.} \tag{5.3}%
\end{equation}
Using these results, we obtain
\begin{align*}
dx  &  =a\alpha r^{\alpha-1}dr\cos\alpha\varphi-ar^{\alpha}\alpha(\sin
\alpha\varphi)d\varphi,\\
dy  &  =a\alpha r^{\alpha}dr\sin\alpha\varphi+ar^{\alpha}\alpha(\cos
\alpha\varphi)d\varphi.
\end{align*}
Based on these results,we arrive at
\begin{equation}
dx^{2}+dy^{2}=a^{2}\alpha^{2}r^{2\left(  \alpha-1\right)  }(dr^{2}%
+r^{2}d\varphi^{2}). \tag{5.4}%
\end{equation}
If we demand that $a^{2}\alpha^{2}=1,$ this then allows us to get rid of $a$
.Next ,we look at Eq.(5.1) and recall that $\left|  dz\right|  ^{2}%
=dx^{2}+dy^{2}=dr^{2}+r^{2}d\varphi^{2}$ and $r=\left|  z\right|  .$ Whence,in
view of Eq.(5.4),we arrive at the result:
\begin{equation}
\alpha-1=\beta. \tag{5.5}%
\end{equation}
Furthermore,if we perform the following rescaling:
\begin{equation}
t=\alpha\varphi\text{ ,}\rho=\alpha^{-1}r^{\alpha}, \tag{5.6}%
\end{equation}
then the metric given by Eq.(5.4) is converted into that given by Eq.(5.2)
with $\rho\leftrightarrows r.$ Clearly,it makes sense to choose $\alpha
=\frac{\theta}{2\pi}$ \quad where $\theta$ is the cone angle.Then,Eq.(5.5)
produces:
\begin{equation}
\theta=2\pi(\beta+1) \tag{5.7}%
\end{equation}
in complete agreement with the result of Troyanov$[10]$ where it was obtained
in a somewhat different way.$\square$

\textbf{Corollary} \textbf{5.2. }\textit{By employing the results of Hopf,e.g.
see Eq.s (4.21)-(4.23), we obtain the index of the quadratic differential:}%

\begin{equation}
I(p_{i})=1-\alpha_{i}=1-\frac{\theta_{i}}{2\pi}=-\beta_{i}\text{ .} \tag{5.8}%
\end{equation}

\textbf{Corollary} \textbf{5.3}.\textit{The Poincare'-Hopf index
theorem,Eq.(1.9),Part I, can now be written at once as }%
\begin{equation}
\chi=\sum\limits_{i}(1-\frac{\theta_{i}}{2\pi}). \tag{5.9}%
\end{equation}
\textit{This result is also in agreement with that obtained in Ref.[10]
where,again,it was obtained differently.}

\mathstrut

Since Troyanov is not using Hopf's arguments explicitly,there is no
restriction on I(p$_{i})$ to be an integer or halfinteger (e.g.see Definition
4.1.)in his work.We also (only for a moment!) will suppress this restriction
in order to discuss some known facts about classical 2+1 gravity .

\subsection{\quad2+1 gravity and quadratic differentials}

In the system of units in which the speed of light c=1,the Einstein's
equations are known to be $[49]$
\begin{equation}
G_{\beta}^{\alpha}=8\pi GT_{\beta}^{\alpha}\text{ \quad; }\alpha,\beta=1-4.
\tag{5.10}%
\end{equation}
Here G is the gravitational constant,$T_{\beta}^{\alpha}$ is the
energy-momentum tensor and $G_{\beta}^{\alpha}$ is the Einstein's tensor,
\begin{equation}
G_{\beta}^{\alpha}=R_{\beta}^{\alpha}-\frac{1}{2}R\delta_{\beta}^{\alpha
}\text{ ,} \tag{5.11}%
\end{equation}
R is the scalar curvature( in the case of 2 dimensions $\frac{R}{2}=K$ where
$K$ is defined in Eq.(4.4)) and $R_{\beta}^{\alpha}$ is the Ricci
tensor(obtained by contraction from the Riemann curvature tensor).In 2+1
dimensions in synchronous system of coordinates $[50]$ the first fundamental
form is given by
\begin{equation}
dl^{2}=-dt^{2}+\gamma_{ij}dx_{i}dx_{j}\text{ , }\mathbf{x=\{}x_{1},x_{2}\}
\tag{5.12}%
\end{equation}
(to be compared with Eq.(4.1)).The Einstein tensor $G_{\beta}^{\alpha}$ has
only one nonzero component $[9]$ $G_{0}^{0}=-\frac{1}{2}R$ and ,accordingly,
$T_{\beta}^{\alpha}$ also has only one nonzero component $T_{0}^{0}$ \quad
given by
\begin{equation}
T_{0}^{0}=-\sum\limits_{i}m_{i}\frac{1}{\sqrt{\gamma}}\delta^{2}%
(\mathbf{x-x}_{i}^{{}})\text{ .} \tag{5.13}%
\end{equation}
Multiplying both sides of Eq.(5.10) by $\sqrt{\gamma}$ (where $\gamma$ is
$\det\gamma_{ij})$ and integrating over the surface we obtain,
\begin{equation}
\frac{1}{8\pi G}\int d^{2}x\sqrt{\gamma}K=\sum\limits_{i}m_{i}\text{ .}
\tag{5.14}%
\end{equation}
Since $\frac{1}{2\pi}\int d^{2}x\sqrt{\gamma}K$ is the Euler characteristic
,$\chi=2-2g$,of the surface of genus $\ g$ ,we conclude ,that \textbf{at least
in }2+1 \textbf{dimensions} ,the integrated Einstein,\ equations
coincide\textbf{\ }with the\textbf{\ }Poincare-Hopf index
theorem,Eq.(1.9),Part I. \textbf{Conversely,one can arrive at correct Einstein
equations starting from the P-H theorem.} In this case,by the way,there is no
need to invoke the equivalence principle as it is traditionaly done
.Accordingly,there is no need to worry about the justification of the Mach
principle \footnote{*According to W.Pauli ,''Theory of
Relativity''(Dover,NY,1981),the Mach principle is a postulate ''that the
inertia of matter is solely determined by the surrounding masses.It must
therefore vanish when all other masses are removed...''(page 179).}* since
Eq.(5.14) is automaticaly in accord with it.This becomes especialy evident in
view of the mass quantization to be discussed below.

By comparing Eqs(5.9) and (5.14) we obtain at once
\begin{equation}
\sum\limits_{i}\left(  1-\frac{\theta_{i}}{2\pi}\right)  =\sum\limits_{i}%
4Gm_{i} \tag{5.15}%
\end{equation}
which produces
\begin{equation}
\alpha_{i}=\frac{\theta_{i}}{2\pi}=1-4Gm_{i}\text{ .} \tag{5.16}%
\end{equation}
This result is in complete accord with the results of Deser, Jackiw and 't
Hooft , Ref.9 ,where somewhat different set of arguments was employed to
arrive at the above equation.Taking into account Lemma 5.1 ,we obtain as well
the following result for the metric:
\begin{equation}
dl^{2}=\prod\limits_{i}\left|  z-z_{i}\right|  ^{2\beta_{i}}\left|  dz\right|
^{2}, \tag{5.17}%
\end{equation}
where 2$\beta_{i}=-8Gm_{i}$ in view of Eqs (5.5) and (5.16).This result is
also in agreement with that obtained in Ref.[9].

Consider now a special case of Eq.(5.14): g=0 .Then Eq.s (5.5) and (5.9)
produce the following constraint on $\beta_{i}$ :
\begin{equation}
\sum\limits_{i}\beta_{i}=-2\text{ .} \tag{5.18}%
\end{equation}
Following Troyanov$[10],$ let us consider the change of variables in
Eq.(5.17): w=z$^{-1}.$ This produces
\[
\left|  dz\right|  ^{2}=\frac{\left|  dw\right|  ^{2}}{\left|  w\right|  ^{4}}%
\]
and
\begin{equation}
\prod\limits_{i}\left|  z-z_{i}\right|  ^{2\beta_{i}}=\prod\limits_{i}\left|
w\right|  ^{-2\beta_{i}}\left|  1-wz_{i}\right|  ^{2\beta_{i}}=\left|
w\right|  ^{4}\prod\limits_{i}\left|  1-wz_{i}\right|  ^{2\beta_{i}}\text{ .}
\tag{5.19}%
\end{equation}
Collecting all terms together ,we obtain for the metric
\begin{equation}
dl^{2}=\prod\limits_{i}\left|  1-wz_{i}\right|  ^{2\beta_{i}}\left|
dw\right|  ^{2}\text{ \quad.} \tag{5.20}%
\end{equation}
This result was obtained with account of the constraint ,Eq.(5.18).Hence,the
metric $dl^{2}$ is regular at infinity,w=0, and maintains the same form as
given in Eq.(5.17).This result should hold under the restriction that
$\alpha_{i}$ cannot become negative in view of Eq.(5.2).This happens to be a
very serious restriction.Indeed,according to Eq.(5.14) for the case of g=0 and
just for one mass $m$ we obtain
\begin{equation}
1=2Gm\text{ .} \tag{5.21}%
\end{equation}
Using this value of mass in Eq.(5.16) we also obtain ,
\begin{equation}
\alpha=1-4G\cdot\frac{1}{2G}=-1\text{ .} \tag{5.21}%
\end{equation}
This is not permissible,however,since it contradicts Eq.(5.2). Moreover,let
$g>0$ in Eq.(5.14),then we obtain
\begin{equation}
\sum\limits_{i}m_{i}\leq0\text{ ,}g>0\text{ .} \tag{5.22}%
\end{equation}
This result is ,apparently ,meaningless as well since we expect our masses to
be nonnegative.

So far ,we have not imposed an additional constraint coming from the Hopf
quantization rule, Definition 4.1.If we constrain our indices to the
(half)integers ,then we obtain the following \textbf{mass}
\textbf{quantization condition} :
\begin{equation}
4Gm_{i}=\left\{
\genfrac{}{}{0pt}{}{1}{\frac{1}{2}}%
\right.  \text{ \quad.} \tag{5.23}%
\end{equation}
In the first case we have the degenerate case ,$\alpha=0,$ according to
Eq.(5.16) ,and in the second ,we obtain , $\alpha=\frac{1}{2}$ which produces
prong-type singularity,e.g. see Fig.2 of Part I which has the index $\frac
{1}{2}.$

So far ,we have not invoked more recent results obtained by Rivin$[11].$If we
use his results,then instead of Eq.(5.9) ,we obtain ($\theta_{i}>0):$%
\begin{equation}
\chi=\sum\limits_{v_{i}\notin\partial S}(1-\frac{\theta_{i}}{2\pi})+\frac
{1}{2}\sum\limits_{v_{j}\in\partial S}(1-\frac{\theta_{i}}{\pi}) \tag{5.24}%
\end{equation}
where the first sum runs ,as before,over the conical singularities while the
second runs over the truncated(punctured ) conical singularities.The above
extension of Troyanov's results in spite of its simple look is highly
nontrivial and is not straightforwardly obtainable.This result becomes
especially useful if we would like to remove the restriction on the total mass
of our ''Universe'',e.g. for g=0 using Eq.(5.15) we obtain ,
\begin{equation}
\frac{1}{2G}=\sum\limits_{i}m_{i}\text{ ,} \tag{5.25}%
\end{equation}
in accord with Ref.[9]To remove this restriction, let us consider ,for
example,how Eq.(5.24) works for a disk $D^{2}$ (evidently , $S^{2}$can be
obtained by gluing two disks) .If the ''boundary sum'' is ignored and only the
nondegenerate cone angles are considered,then, in agreement with the results
of Part I ,we should have 2 thorn-like singularities using the quantizaton
condition (5.23) and the fact that for the disc $\chi(D^{2})=1.$ We can place
yet another thorn on $D^{2}$ if we use the ''boundary sum''term.Without
violating the requirement $\theta_{i}>0$ and requiring $\frac{\theta_{i}}{\pi
}$ to be a positive integer$[13],$ we obtain the first nontrivial result for
$\frac{\theta_{i}}{\pi}=2$ which produces the desired -$\frac{1}{2}$ factor
characteristic of Y-type defects (Fig.2,Part I).The rest of the arguments used
in Part I now can go through without change so that the restriction for the
total mass can now be removed. Evidently,the extension of these results to
higher genus surfaces becomes also possible without any problems.

\textbf{Remark} \textbf{5.4}. \textit{At this point it is appropriate to
remind the reader that even for the Coulombic charges on S}$^{2}$\textit{ the
topology requires only two charges in order for the P-H theorem to be
satisfied(e.g.see section 1 ,Part I).This is completely analogous to the total
mass restriction ,Eq.(5.25), for the case of gravity.The electroneutrality
requirement emerges naturally if we want to put the additional charges on
S}$^{2}$\textit{ .Moreover,although we actually can put only the
''sources''(''+'') and the ''sinks''(''-'') on S}$^{2}$\textit{ ,the presence
of these additional singularities (both having index +1) automatically creates
the induced \quad saddles (with index -1).Something similar occurs in the case
of gravity because of an extra boundary sum term (''Rivin sum'' term).}

\textbf{Remark} \textbf{5.5}.\textit{The presence of such boundary term(s)
becomes obvious if one wants to consider Eq.(5.14) for surfaces of genus g
higher (or equal) than one.Since for g=1 the l.h.s. of (5.14) is zero,we
obtain :}$\sum\limits_{i}m_{i}=0$\textit{ (''electroneutrality'') which is of
limited use (photons,neutrinos,etc)for gravity.For }$g>1$\textit{ we obtain
}$\sum\limits_{i}m_{i}<0$\textit{ and this is physically problematic
(please,see Note added in proof at the end of this paper).}

\textbf{Remark 5.6}.\textit{In the original work by Troyanov}$[10]$%
\textit{there is no restriction on }$\alpha_{i}$\textit{ to be in the range
between 0 and 1.It can be any positive integer or halfinteger(Definition
4.1).Under such conditions the relation given by Eq.(5.9) becomes correct for
surfaces of any genus.Unfortunately,it cannot be used for gravity for reasons
explained above.}

\textbf{Reamark} \textbf{5.7}.\textit{It is useful to recall why altogether
one should be concerned with surfaces of higher genus in the case of 2+1
gravity.Following Petrov}$^{51},\operatorname{Ei}$\textit{nstein spaces are
characterized by the condition: }%
\begin{equation}
R_{ij}=\tilde{\lambda}g_{ij}\text{ .} \tag{5.26}%
\end{equation}
\textit{Since the scalar curvature }$R=R_{i}^{i}=g^{ik}R_{ik}$\textit{ ,the
constant }$\tilde{\lambda}$\textit{ in Eq.(5.26) can be eliminated with the
result: }%
\begin{equation}
R_{ij}=\frac{R}{d}g_{ij}, \tag{5.27}%
\end{equation}
\textit{where d is the dimensionality of space-time.Using this result
,Einstein tensor,Eq.(5.11), can be rewritten as }%
\begin{equation}
G_{\beta}^{\alpha}=\left(  \frac{1}{d}-\frac{1}{2}\right)  \delta_{\beta
}^{\alpha}R \tag{5.28}%
\end{equation}
\textit{so that the covariant derivative of both sides of eq.(5.10) produces }%
\begin{equation}
G_{\beta,\gamma}^{\alpha}=0\text{ .} \tag{5.29}%
\end{equation}
\textit{Since }$T_{\beta,\gamma}^{\alpha}=0$\textit{ by construction,Eq.(5.29)
can be equivalently rewritten (with account of Eq.(5.28) ) as }%
\begin{equation}
\left(  \frac{1}{d}-\frac{1}{2}\right)  R_{,\gamma}=0\text{ .} \tag{5.30}%
\end{equation}
\textit{For }$d\neq2$\textit{ we have to study spaces of constant curvature
called Einsein spaces.For d=2 (that is for the fixed time slice) we have no
choice but to conclude that any two dimensional surface is Einstein space.But
any two dimensional surface is the Riemann surface }$^{46}.$

\section{Some more advanced topics:a brief discussion}

\subsection{ Inclusion of the cosmological term}

Einstein's Eq.(5.10) is written without the cosmological constant $\Lambda$
term.This deficiency can be easily corrected.By multiplying both sides of
Eq.(5.10) by $\sqrt{\gamma}$ and taking into account that$[9]$
\begin{equation}
-\sqrt{\gamma}G_{0}^{0}=\frac{1}{2}\sqrt{\gamma}R \tag{6.1}%
\end{equation}
and $R/2=K$ (Eq.(4.4)) where,for the metric given by $dl^{2}=\rho\left|
dz\right|  ^{2},$ the Gauss curvature $K$ is known to be$[52]$%
\begin{equation}
K=-\frac{2}{\rho}\frac{\partial^{2}\ln\rho}{\partial z\partial\bar{z}}\text{
\quad.} \tag{6.2}%
\end{equation}
Using these results along with Eq.s(5.10) and (6.2) we obtain,
\begin{equation}
\frac{\partial^{2}\varphi}{\partial z\partial\bar{z}}+\frac{\Lambda}{2}%
\exp\varphi=-8\pi G\sum\limits_{i}m_{i}\delta^{2}(z-z_{i}) \tag{6.3}%
\end{equation}
where $\rho=\exp\varphi$ .For $\Lambda=-1$ and $8Gm_{i}=1$ the above equation
coincides with the Liouville equation discussed by Takhtadjian$[14]$ in
connection with the nonperturbative approach to string
theory.Accordingly,quantization of 2+1 gravity and 1+1 string theory are
related to each other as it was noticed already by Witten$[53].$Consider now
the Liouville equation in the punctured complex plane that is in \textbf{C}%
$\backslash$%
$\{z_{1},...,z_{n}\},$and let $\Lambda$ be some \textbf{positive
}constant.This case (actually corresponding to the punctured sphere S$^{2})$
was treated in Ref.54. The authors of Ref.54 were able to prove the following

\textbf{Theorem} \textbf{6.1}.\textit{The Liouville equation }%
$\bigtriangledown^{2}\varphi=-\exp(2\varphi)$\textit{ in the punctured complex
plane }$+\{\infty\}$\textit{ has solution.Near each puncture the solution is
represented by }$\varphi(z)=\beta_{i}\ln\left|  z-z_{i}\right|  +$%
\textit{harmonic function.The constants }$\beta_{i}=\frac{\theta_{i}}{2\pi}%
-1$\textit{ and satisfy the restriction: }%
\begin{equation}
\sum\limits_{i}\beta_{i}\geq-2 \tag{6.4}%
\end{equation}
\textit{and}
\begin{equation}
\left(  \sum\limits_{i\neq j}\beta_{i}\right)  -\beta_{j}<0\text{ for all
}j=1,...,n. \tag{6.5}%
\end{equation}

\textbf{Corollary} \textbf{6.2.}From the above theorem it follows ,that the
results of Troyanov$[10]$ (and Rivin$[11])$ In the light of the results
presented in the previous section, it appears ,that a similar proof can be
oallow us to by-pass solution of the Liouville Eq. (6.3) (at least for the
case of S$^{2}).$

In the light of the results obtainedin the previous section,it appears,that a
similar proof can be obtained for $\Lambda<0$ as well.This is,however,not the
case due to the results of McOwen$[55]$ .His results are analogous to that
discussed by Takhtadjian$[14].$Moreover,the same results were obtained much
earlier by Nevanlina(e.g.see pages 249-250 of Ref.[56] and take into account
the very minor typos,e.g.sign errors,etc).The difference between treatments of
$\Lambda>0$ and $\Lambda<0$ cases deserves further study in the light of
Troyanov's results

\subsection{ Some sum rules}

Although the Poincare-Hopf index theorem plays the major role in study of
topology of two dimensional surfaces,sometimes,additional ramifications are
required. Let us begin with

\textbf{Lemma} \textbf{6.3}.\textit{If the quadratic differential has poles
with the total order p and zeros (including their multiplicity) of total order
q,then for the Riemann surface of genus g with n boundary components we must have}%

\begin{equation}
p-q=4-4g-2n. \tag{6.6}%
\end{equation}

\textbf{Proof}.Consider first the case of n=0.Then,taking into account
Eqs(4.21)-(4.23)and using the P-H theorem,Eq.(1.9),part I,we obtain,
\[
\sum\limits_{poles}n_{i}-\sum\limits_{zeros}n_{j}=2(2-2g)\text{ \quad.}%
\]
Let now we have n boundary components.These can be obtained in the following
way.Consider a sphere S$^{2}$ with g handles and add one additional handle so
that the genus of the surface becomes g+1.Let us now squeeze the additional
handle so that it breaks into 2 pieces leaving 2 punctures on S$^{2}.$Hence,we
obtain the correspondence: $g\leftrightarrows2n$ .According to Jenkins$[57]$
(see also Rivin $[11])$ the \textbf{boundary} zeros are necessarily of
\textbf{even order} .Therefore,instead of having $2(2-2g)-4n$ we obtain the
result (6.6).$\square$

Because the quadratic differentials are transforming like tensors of rank 2
,e.g.see Eq.(3.2),one may ask a question: under what conditions they can be
formed out of the product of the abelian differentials? Following Ref.[13]
,for the Riemann surface $R$ ,let $k_{i}$ be the order of zero at some
$p_{i}\in R$ and let ,for example,$k_{i}=-1$ if at $p_{i}$ we have a simple
pole(a prong-like singularity),etc ,then each quadratic differential $\phi$ is
associated with the set of data ($k_{1},...,k_{n};\varepsilon)$ where
$\varepsilon=+1$ if $\phi$ is the square of the abelian differential and
$\varepsilon=-1$ if it is not.One can prove the following

\textbf{Theorem} \textbf{6.4}.\textit{Let }$k=(k_{1},...,k_{n};\varepsilon
)$\textit{ where }$k_{i}\in\{-1,1,2,...\}$\textit{ and }$\varepsilon=\pm
1.$\textit{ Then,for a closed surface R there is }$\phi$\textit{ realizing
}$k$\textit{ if and only if}

\textit{\quad a)}$\sigma(k)=0$\textit{ \quad}$\operatorname{mod}%
4,\sigma(k)\geq-4$\textit{ and }$\sigma(k)=\sum\limits_{i=1}^{n}k_{i}$\textit{ ;}

\textit{\quad b)}$\varepsilon=-1$\textit{ if any }$k_{i\text{ }}$\textit{ is
odd and}

\quad c)($k_{1},...,k_{n};\varepsilon)\neq(4;-1),(1,3;-1),(-1,1;-1)$ or (
\quad;-1).

\textbf{Proof}.Please,consult Ref.13.$\square$

In the case if we are interested under what conditions the pseudo-Anosov
regime of surface homeomorphisms is possible,the following theorem could be
also of major importance.To formulate this theorem we have to provide the
following definition first.

\textbf{Definition} \textbf{6.5}.\textit{The 1-prong is thorn-like ( a simple
pole) singularity,the 3-prong is Y-type (a simple zero)singularity .The
p-prong is the singularity with p arms}$^{24}$\textit{ .The data set could be
realized in terms of the prong numbers }$p_{i}:(p_{1},...,p_{n};\varepsilon).$

\textbf{Theorem} \textbf{6.6}.\textit{There is a pseudo-Anosov homeomorphism
on }$R$\textit{ of genus g and n punctures realizing the data }$(p_{1}%
,...,p_{j};\varepsilon)$\textit{ if and only if}

\textit{\quad a) }$\sum\limits_{i=1}^{j}(p_{i}-2)=4(g-1),$

\textit{\quad b)}$\varepsilon=-1$\textit{ if any }$p_{i}$\textit{ is odd,}

\textit{\quad c)}$(p_{1},...,p_{j};\varepsilon)\neq(6;-1),(3,5;-1),(1,3;-1)$%
\textit{ or ( ;-1).}

\textit{\quad d) the number of indices }$i$\textit{ for which }$p_{i}%
=1$\textit{ is less than or equal to }$n$\textit{ (that is}

\textit{\quad the thorns may be ''sitting'' at the punctures).}

\textbf{Proof}.Please,consult Ref.[13].$\square$

\textbf{Remark} \textbf{6.7.}\textit{ Theorems 6.2 and 6.3 provide some
additional selection rules which should be taken into account when the mass
spectrum of 2+1 gravity is of interest.They do not follow trivialy from the
P-H theorem.}

\subsection{ Connections with the theory of Teichm\"{u}ller spaces}

In this subsection we would like to discuss briefly under what
\textbf{physical }conditions it is possible to anticipate the transitions from
the periodic to the pseudo-Anosov regime.These results are complementary to
that presented in section 5 of Part I.

We have encountered a glimpse of the Teichm\"{u}ller theory in section
4.2.From it,we realize that we are in the domain of the Teichm\"{u}ller theory
as soon as instead of the conformal mapping (for which $\Phi_{\bar{w}}=0)$ we
are dealing with the quasiconformal (for which $\Phi_{\bar{w}}\neq0).$ The
length element given by Eq.(4.1) can be brought into form$[58]$%
\begin{equation}
dl^{2}=\Lambda(z,\bar{z})\left|  dz+\mu(z)d\bar{z}\right|  ^{2} \tag{6.7}%
\end{equation}
where
\[
\Lambda=\frac{1}{4}(E+G+2\sqrt{EG-F^{2}})
\]
and
\[
\mu=\frac{E-G+2iF}{E+G+2\sqrt{EG-F^{2}}}\text{ \quad}%
\]
in notations of section 4.1.At the same time,the length element given by
Eq.(4.1) can be brought into the conformal form given by Eq.(4.6).Using
Eq.(4.6) ,let $w=x_{1}+ix_{2}\equiv u+iv$ .Then, we have as well ,
\begin{equation}
dl^{2}=\rho\left|  dw\right|  ^{2}=\rho(du^{2}+dv^{2}). \tag{6.8}%
\end{equation}
If now $w=w(z,\bar{z}),$ we obtain ,
\begin{equation}
\left|  dw\right|  ^{2}=\left|  w_{z}\right|  ^{2}\left|  dz+\frac{w_{\bar{z}%
}}{w_{z}}d\bar{z}\right|  ^{2}, \tag{6.9}%
\end{equation}
where ,as before,$w_{z}=\frac{\partial}{\partial z}w$ ,etc .Comparison between
Eqs.(6.7)-(6.9) produces:
\begin{equation}
\Lambda(z,\bar{z})=\rho\left|  w_{z}\right|  ^{2}. \tag{6.10}%
\end{equation}
This result coincides with Eq.(3.2) as required.Comparison between Eqs.(6.7)
and (6.9) produces as well
\begin{equation}
\mu(z)=\frac{w_{\bar{z}}}{w_{z}}\text{ .} \tag{6.11}%
\end{equation}
This is known as \textbf{Beltrami }equation. If $w_{\bar{z}}=0,$ then $\mu=0$
and the mapping is conformal.Looking at Eq.(4.35) we can write as well $[29]$%
\begin{equation}
\left|  \mu(z)\right|  \leq\frac{\hat{K}-1}{\hat{K}+1}<1. \tag{6.12}%
\end{equation}
Either$\left|  \mu\right|  $ or $\hat{K}$ determine the \textbf{amount of
stretching }the surface undergoes(this strething can include twisting as well).

Let us now introduce notations:
\[
\mu(z)\leftrightarrows\mu_{w}(z)\text{ \quad if }\mu\text{ is defined through
Eq.(6.11),}
\]
\begin{equation}
K_{w}=\frac{1+\left|  \mu_{w}(z)\right|  }{1-\left|  \mu_{w}(z)\right|
}=\frac{1+k}{1-k}\text{ , }0<k<1. \tag{6.13}%
\end{equation}
Using these notations one can prove the following

\textbf{Theorem} \textbf{6.8}. \textit{Let }$\varphi$\textit{ be some
quadratuc differential on }$R$\textit{ ,then}
\begin{equation}
\mu_{w}(z)=k\frac{\bar{\varphi}(z)}{\left|  \varphi(z)\right|  } \tag{6.14}%
\end{equation}
\textit{where }$\bar{\varphi}$\textit{ means the complex conjugate}.

\textbf{Proof.} Please, consult Ref.[59].$\square$

\bigskip\textbf{Definition} \textbf{6.9}. \textit{A diffeomorphism (posibly
with isolated singularities) }$f:R\longrightarrow R^{\prime}$\textit{ is
admissible if}
\begin{equation}
K_{f}\equiv K[f]<\infty\text{ .} \tag{6.15}%
\end{equation}

\textbf{Theorem} \textbf{6.10}.\textit{In order for diffeomorphism to be
admissible ,we have to require}
\begin{equation}
K[f\circ f]=[K[f]]^{2}\text{ .} \tag{6.16}%
\end{equation}

\textbf{Proof}.Please,consult Ref.[60].$\square$

\bigskip\textbf{Corollary} \textbf{6.11}.\textit{Using Theorem 6.4. it is
straightforward to show that if }$\left|  \mu_{f}(z)\right|  =k$\textit{ (see
Eq.(6.14)) ,then }$\left|  \mu_{f\circ f}(z)\right|  =2k/(1+k^{2}).$\textit{
By combining Eq.s (6.13)-(6.16) we observe that if }$K[f]$\textit{ was
initialy }$>1,$\textit{then}
\[
K[f\circ f]=[K[f]]^{2}>K[f]>1.
\]

\quad Whence, successive dilatations lead to the successive stretching.This
fact leads to the most profound implications reflected in the following
theorem and corollaries that follow.

\textbf{Theorem} \textbf{6.12}.\textit{Let }$\varphi$\textit{ be associated
with the admissible map }$f$\textit{ of }$R$\textit{ . If }$\varphi$\textit{
has only a finite number of critical points ,then }$\varphi$\textit{ has no
saddle connections (critical trajectories running between these critical
points) . In particular,every trajectory of }$\varphi$\textit{ is dense in
}$R$\textit{ }

\textbf{Proof}.Assume that after some iteration $\quad f^{m}=\underbrace
{f\circ\cdot\cdot\cdot\circ f}_{m}$ we had been able to fix all saddle
connections.Then,on one hand,the length between them should remain the same as
it was initially and,on another hand,in view of Eq.(6.16) the length has
expanded by the factor of $K^{\frac{m}{2}}>1.$This contradiction proves the
theorem . Consequently ,every trajectory of $\varphi$ is dense in $R$
(otherwise ,there would be saddle connections).$\square$

\textbf{Corollary} \textbf{6.13.a}) \textit{The theorem 6.5. indicates that
the existence of the meandritic labyrinths ,discussed in Part I, is possible
only if the surface undergoes some stretching.Hence , the stretching is needed
to destroy the reducible and periodic phases and to create the pseudo-Anosov phase.}

\textbf{Corollary} \textbf{6.13.b})\textit{If the Riemann surface }$R$\textit{
has some boundaries which are\ pointwise fixed,then there is no admissible
mapping in any homotopy class of self-mappings.That is the non-slip boundary
conditions must be violated}$[61,62]$\textit{ in order for the pseudo-Anosov
phase to become possible.(This is in complete agreement with Thurston's
''earthquakes''mentioned in Part I).}

\textbf{Corollary} \textbf{6.13.c})\textit{The deformations of }$R$\textit{
may take place in real time .The evolution of such surface is described by the
product }$R\times\lbrack0,t]$\textit{ which is just some 3-manifold (see next
subsection).This type of manifold was suggested by Witten in connection with
2+1 gravity}$[53]$\textit{ (without matter fields).Inclusion of the matter
fields produces the pseudo-Anosov foliations evolving in real time.This can be
most conveniently explained in terms of braids through connection with the
meandritic labyrinths discussed in Part I.We provide some details related to
this connection in section 7.5.In addition,we would like to describe briefly
the emerging connections with 3-manifolds .This is accomplished in the
remainder of this paper and in the Appendix.}

\subsection{ 2+1 gravity and hyperbolic 3-manifolds}

In this subsection and in the Appendix,without any pretense on completeness,we
would like to mention several facts related to 3-manifolds.More detailed
analysis would require a separate publication(s) and is left for the future .

There are several ways to construct 3-manifolds.For the comprehensive
treatment of this subject,e.g. see Refs.[63,64].A quick introduction into this
field is beautifully presented in Refs[36] and [65].In particular,following
Ref.[65], we would like to describe one of the methods of constructing
3-manifolds which is naturally connected with the results of the preceding subsection.

Let $S$ be a compact surface (2 manifold) ,perhaps,with boundary.Let $I$=[0,1]
be the closed interval of real numbers from 0 to 1.Then,the Cartesian product
$S\times I$ is 3-manifold with boundary.Consider now the bottom,i.e.$S\times
\{0\},$ and the top,i.e.$S\times\{1\},$ surfaces and let us glue them
together.The operation of gluing is trivial if the top surface is the same as
the bottom surface(that is if the bottom surface was parallel translated to
the top).But,the situation becomes more intresting if one considers the
surface homeomorphisms $h:S\longrightarrow S^{\prime}.$These homeomorphisms
are superimposed with the gluing operation in which each point $(x,0)\in
S\times\{0\}$ of the bottom surface is identified with the corresponding point
$(f(x),1)\in S\times\{1\}$ of the top surface.The result of such an
identification is new 3-manifold which is actually a fiber bundle over the
circle $S^{1}$ (compare this discussion with that of section 3,Part I )each of
the fibers being the original surface at given instant of ''time''.The fact
that thus constructed manifold is a fiber bundle is in accord with the
discussion presented in Thurston's book (Ref.[36],page 159) where some
additional details could be found.

Although we just had explained the construction of a typical 3-manifold,it is
still unclear from this construction why this manifold has to be hyperbolic
and why 2+1 gravity should be related to the hyperbolic 3-manifolds(as well as
to knots and links).The connection between 3-manifolds and knots and links was
established by Lickorish $[66]$and Wallace $[67]$ who proved the following

\textbf{Theorem} \textbf{6.14}\textit{.Every closed ,connected ,orientable
3-manifold is obtained from }$S^{3}$\textit{ by removing mutually exclusive
(but ,perhaps,knotted and linked)collection of solid(framed) tori \{T}$_{i}%
\}$\textit{ ,i=1-n,and then sewing them back into }$S^{3}$\textit{ in a
different way.}

From the Theorem 6.6. it is not immediately clear how 3-manifolds constructed
with help of the fiber bundle methods are related to that constructed by Dehn
surgery methods used in the theorem by Lickorish and Wallace.Leaving the
answer to this question aside,e.g.see Ref.[64] for more details,we arrive at
the following

\textbf{Corollary} \textbf{6.15}.\textit{Every closed ,connected,orientable
3-manifold is a complement of some knot or link.}

Having this Corollary, the question arises:under what conditions the dynamics
of 2+1 gravity and /or dynamics of liquid crystals can be associated with
knots/links or,alternatively ,with 3-manifolds (obtained with help of Dehn
surgery or fiber bundle methods)?

In 1970 Geroch$[68]$ had proven the following

\textbf{Theorem} \textbf{6.16}.\textit{If the open set N is globaly
hyperbolic,then if it is considered as a manifold,it is homeomorphic to }%
$\sum^{(3)}\times R$\textit{ where }$\sum^{(3)}$\textit{ is some 3-dimensional
manifold and \quad}$\forall a\in R$\textit{ the product }$\sum^{(3)}%
\times\{a\}$\textit{is the Cauchy surface .}

\textbf{Remark 6.17}\textit{.For the precise definition of the Cauchy
surface,please,consult Ref.[69]}.

\textbf{Remark} \textbf{6.18.}\textit{Theorem 6.16. admits natural extension
to the case of 2+1 gravity.In this case,we have to replace }$\sum^{(3)}%
$\textit{ with }$\sum^{(2)}$\textit{ which is the Riemann surface of some
finite genus.This is in complete accord with the results of Witten}$[53].$

The products $\sum^{(2)}\times\mathbf{R}$ had been recently discussed in
Ref.[70] (and references therein).In Ref.[70] no attempts were made towards
physical applications .Below ,and ,in part,in the Appendix we shall provide
some condensed summary of the results mainly associated with Ref.[70],which
could be associated with dynamics of 2+1 gravity.

Let us return to the Theorem 6.7 in order to explain the meaning of the words
''globally hyperbolic''.From Part I we know already about the hyperbolic
Poincare disc $D^{2}$ and the hyperbolic half plane H-model(section 4).These
results can be easily extended.For example,the hyperbolic $H^{3}$ model can be
defined by analogy with Eq.(4.1) (Part I):%

\begin{equation}
H^{3}=\{(x,y,t)\in\mathbf{R}^{3},z=x+iy=\mathbf{C,}t>0\} \tag{6.17}%
\end{equation}
and,instead of the open disc $D^{2}$ model defined by Eq.(4.2),now we have to
consider an open ball $B^{3}$with sphere $S_{\infty}^{2}$ at infinity
replacing the circle $S_{\infty}^{1}$ at infinity for $D^{2}$. Instead of the
Fuchsian groups SL(2,\textbf{R)/}$\mathbf{\{}\pm I\},$ we need to use now the
quasi-Fuchsian (or Kleinian) groups $\Gamma$ generated by SL(2,\textbf{C)/}%
$\mathbf{\{}\pm I\}$ .

\textbf{Definition} \textbf{6.19}. \textit{A hyperbolic 3-manifold }%
$M$\textit{ is a quotient }$M=H^{3}/\Gamma$\textit{ of the hyperbolic }$H^{3}%
$\textit{ space by the Kleinian group }$\Gamma$\textit{ .}

\textbf{Theorem} \textbf{6.20}. \textit{Let }$h:S\longrightarrow S$\textit{ be
the pseudo-Anosov homeomorphism.Then, the 3-manifold obtained by fibering over
the circle }$S^{1}$\textit{is hyperbolic and has finite( hyperbolic) volume.}

\textbf{Proof}.Please,consult Ref.[70].$\square$

\textbf{Corollary} \textbf{6.21.a})\textit{Since the volume of such designed
3-manifold is finite,it is associated with knots(links).That is such manifold
is a complement(in }$S^{3})$\textit{ of some knot or link(Theorem 6.6).}

\textbf{Proof}. Please,consult Ref.[36].$\square$

\textbf{Corollary} \textbf{6.21.b})\textit{ If we take the hyperbolic
3-manifold that fibers over the circle, with fiber being a 2d surface ,and
unwrap the covering space by unwrapping the circle direction,then the volume
of thus obtained manifold is infinite.}

\textbf{Proof}.Please,consult Ref.[36] (page 258).$\square$

\textbf{Remark} \textbf{6.22}.\textit{The result just presented is in complete
accord with the conclusion made in Ref.[69](e.g.see paragraph 6.4).That is ,
the physical space-time is noncompact and is a universal covering space of the
space-time which is compact.To prove the existence of such compactified space
is highly nontrivial.Accordingly, the existence of knots and links in 2+1
gravity is not selfobvious.}

\textbf{Remark 6.23}.\textit{Theorem 6.20. is in complete accord with the
Theorem 6.12.discussed earlier.That is ,the surface motion which is associated
with stretching ,which takes place in real time ,creates the hyperbolic
3-manifolds ( provided that the issue of compactification is resolved).}

\textbf{Remark} \textbf{6.24.}\textit{The previous remark allows us to provide
, yet another, interpretation of the process of compactification through the
notion of the universal Teichm\"{u}ller curve which was introduced by
Bers}$[71]$\textit{ and was discussed in physical literature in connection
with problems related to string theories }$[72]$\textit{.This and other topics
related to compactification are summarized in the Appendix.}

\section{Discussion}

\subsection{ Quantization of 2+1 gravity}

The results presented in Parts I and II provide mainly classical description
of 2+1 gravity.The quantum description may require computations which are
based on the master equation, Eq.(4.13) ,of Part I. That is one has to
consider some sort of the random walk on the mapping class group of some 2
dimensional surface of genus $g>1,$possibly with punctures.To make these
computations meaningful it is essential to find physically unambiguous way by
which the transition amplitudes $W_{ij}$ in Eq.(4.13) of Part I can be
calculated.To this purpose,we anticipate ,that the physical results obtained
by 't Hooft ,e.g.see Refs.[73-76], and their development by Franzosi and
Guadagnini$[77]$ could be very helpful. In addition,the results of Appendix
suggest that the quantization could be achieved as well by methods of non
commutative geometry applied directly to the Teichm\"{u}ller space(more
correctly,to the Bers slice).According to Ref.[78],''The qiuasiconformal
charts provide enough analysis to ''quantize the manifold'' in the sense of
constructing a Hilbert space and relevant operator replacing curvature... The
involved Hilbert space theoretical data are of the same nature as those
appearing in the transfer matrix theory of statistical mechanics and suggest a
purely combinatorial approach...in the extended context of spaces with
singularities''. In the physical language ,the transfer matrix approach
discussed in Ref.[78] can be easily recognized as Feynman-Wiener-Kac -type of
path integral calculations.

\subsection{Connections with the theory of dislocations and disclinations}

Connections between 2+1 gravity and the theory of defects and textures in
solids was recently discussed by Katanaev and Volovich,Ref.[15] (Part I).These
authors had noticed that the theory of wedge and edge dislocations as well as
the linear disclinations in solids is \textbf{isomorphic }to the theory of 2+1
gravity of Deser, Jackiw and 't Hooft. Using arguments which are completely
different from those presented in sections 4-6 of Part II ,Katanaev and
Volovich nevertheless had arrived at correct mass (Hopf) quantization
condition, Definition 4.1..They missed ,however,the sum rules of our section 6
which could be obtained only with help of the theory of quadratic differentials.

\subsection{Connections with the theory of motions in classical and quantum billiards}

The problem of quantization of 2+1 gravity (as well as motion of point-like
defects in the presence of wedge and edge dislocations,etc) could be
considered also from the point of view of classical and quantum motion of
particles in the billiards.The connection between the quadratic differentials
and billiards is known to matematicians for some time.In this regard reading
of the encyclopedic references, Refs[79-81],is most helpful.Reading of the
related literature,e.g. Refs [61] and[82] as well as Ref.[83] ,may be helpful
as well.Since in the mathematical physics literature currently there is a
strong interest in the detailed study of various mesoscopic systems$[84],$ we
anticipate that some ''cross fertilization'' between different domains of the
same discipline (physics) could produce some unexpected results.

\subsection{Crumpling,fracture,brittleness}

In his wonderful book , Ref.[36],Thurston discusses the construction of what
he calls a ''hyperbolic paper'' . This construction is based on the famous
theorem by Hilbert

\textbf{Theorem} \textbf{7.1}.\textit{There is no complete smooth surface
embedded in Euclidean three-space with the local intrinsic geometry of
pseudosphere (e.g.see the ''beak of the bird '' in Fig.6)}

That is when the surface crumples,it necessarily develops the conical-like
''beaks''.The mathematics of this crumpling process is discussed in Thurston's
famous lecture notes (e.g.see Ref.[20] , Part I)and,surely,involves the train
tracks.At the same time,in physics literature the process of crumpling of
surfaces was recently discussed in Refs.[85,86].The process of crumpling is
closely related to brittleness and fracture.Depending upon the rigidity of
surface,it may or may not ''want'' to crumple under the applied stresses.In
the last case case,we may observe some cracks(more exactly,the crack
patterns,e.g.see Ref.87, which eventually cause the disintegration of the
surface.The dynamical equations for such crack patterns are similar to the
master equation, Eq.(4.13), Part I,for 2+1 gravity.Evidently,one can get some
additional insights into both fields if one is aware about the existence of
the other.Since microscopically the cracks are caused by
dislocations/disclinations,e.g.see section 7.2.,surely,one can think about the
dynamics of 2+1 gravity in terms of the dynamics and topology$[88]$ of
fracture $[89]$ .

\subsection{Meanders,the Temperley-Lieb algebra and the invariants of 3-manifolds}

In Part I,section 5,Fig.25,we have mentioned that any meander (or disconnected
system of meanders)can be built by superposition of two arc configurations of
the same order: one is being considered as the top while another as the
bottom.In Ref.[36],Part I,it is shown,that \textbf{both }the top and the
bottom configurations can be obtained with help of the product of the
Temperley-Lieb (TL) algebra generators(since these generators admit braid-like
graphical representation) $\{e_{i}\}$ which obey the TL algebra TL$_{n}%
$($\delta)$ given by :%

\[
e_{i}e_{j}=e_{j}e_{i}\text{ \quad if }\left|  \text{i-j}\right|  >1
\]%

\[
e_{i}^{2}=\delta e_{i\text{ }}\text{ , \quad}i=1,...,m-1
\]%

\[
e_{i}e_{i\pm1}e_{i}\text{ =}e_{i}\text{ , }i=1,...,m-1
\]
where $\delta$ is some constant.For the TL algebra composed of n
elements,1,$e_{1},...,e_{n-1}$ ,the number of possible \textbf{independent
}products of $e_{i}$ generators is given by the Catalan number,Eq.(3.9),Part
I. Following Lickorish$[90,91]$ ,one can think of the space of these
independent products as a vector space $V_{n\text{ }}$ of dimension $C_{n}$
.Then,one can construct a bilinear form
\[
<\text{ },\text{ }>\text{ }:\text{ }V_{n}\times V_{n}%
\]
which in ordinary language means just a kind of a pairing between the top and
the bottom arc configurations as discussed above.As it was shown by
Lickorish,the above bilinear form plays the central role in constructing the
algebraic invariants of 3-manifolds.Hence,use of the meanders allows to
provide the alternative (to the Chern-Simons$[92])$ formulation of the
invariants of 3-manifolds.Additional connections between the TL algebra and
the meanders could be found based on the notion of parenthesis$[92]$.The
parenthesis are also associated with TL algebra.By definition,a parenthesis
diagram is a word in the alphabet with three letters $\bullet$ \textbf{(} and
\textbf{) , }e.g.see page 543 of Ref.[91].Exactly the same three letter
alphabet is used for the description of meanders,e.g.see page 121 of Ref.[34]
(Part I).We hope,that these connections deserve further study.

$\mathbf{\quad\quad\quad}$

\begin{description}
\item \textbf{Note added in proof.} After this work was completed ,we run
across the only one article by Einstein written for the Scientific American
(Sci.Am.\textbf{182}(4),pages 13-17 (1950))in which he wrote ,e.g. see page 16
(bottom),''The fact that the masses appear as singularities indicates that
these masses themselves cannot be explained by symmetrical $g_{ik}$%
\textbf{\ }fields,or ''gravitational fields''.Not even the fact that only
\textbf{positive }gravitating masses exist can be deduced from the
theory.Evidently a complete relativistic field theory must be based on a field
of more complex nature...''.In the above, the word ''positive'' was emphasized
by Einstein himself. We were mystified by the fact that the theory of
quadratic differentials had begun its development just about the time when
Einstein wrote these words.

\item \textbf{Acknowledgments.}. The author would like to thank Professor
Louis Kauffman (U of Illinois at Chicago) for his constant attention to this
work. The author would also like to thank Professor Klaus Johannson (U of
Tennessee at Knoxville) for his careful reading of the draft of this paper and
for many helpful remarks related to 3-manifolds.Helpful conversations and
correspondence with Professor John Smillie (Cornell U )are also highly
appeciated. Last but not the least,the author had greatly benefited from his
interactions with Professor Dale Rolfsen (U of British
Columbia,Vancouver,Canada).Without Dale's influence the author would not be
aware of many beautiful mathematical results,especially related to topology
and knot theory.
\end{description}

\mathstrut\newpage

\quad\quad\quad\quad\quad\quad\quad\quad\quad\quad\quad\quad

\quad\quad\quad\quad\quad\quad\quad\quad\quad\quad\quad\quad\quad\textbf{Appendix}

In this Appendix we would like to provide an outline of the results recently
obtained in mathematical literature which are related to the problem of
compactification discussed in section 6.4.The purpose of such presentation
also lies in demonstration of nontriviality of the problem of knots/links
relevance to 2+1 gravity.

Let us begin with the following observations.The group of Mobius
transformations $\Gamma$ with \textbf{real }coefficients , PSL(2,R), is
defined through the transformation law
\begin{equation}
\gamma(z)=\frac{az+b}{cz+d}\text{ , \quad\quad}ad-bc=1. \tag{A.1}%
\end{equation}
The discontinuous subgroups (e.g.see Definition A.4. below) of $\Gamma$ are
called Fuchsian groups.These subgroups are classified as \textbf{elliptic
}(respectively ,\textbf{parabolic }and \textbf{hyperbolic }) if $\left|
a+d\right|  <2$ (respectively, $\left|  a+d\right|  =2$ or $\left|
a+d\right|  >2$ ).Compare with Appendix to Part I.

A parabolic transformation is conjugate in $\Gamma$ to the translation:
$z\rightarrow z+1,$ a hyperbolic transformation is conjugate to :$z\rightarrow
\lambda z$ for some $\lambda>1.$ Let $G\in\Gamma$ and let $\pi:$ $H\rightarrow
H/G$ be some Riemann surface,then one can prove the following

\textbf{Theorem} \textbf{A.1}. \textit{If }$H/G$\textit{ is compact ,then each
}$\gamma\in G\backslash\{id\}$\textit{ is hyperbolic}

\textbf{Proof}. Please,consult Ref.[29].$\square$ \quad

\quad These results can be generalized to higher dimensions.E.g.instead of $H
$ we can use $H^{3}$ defined by Eq.(6.17),instead of PSL(2,R) we have to use
PSL(2,C).Theorem A.1. is extendable to $H^{3}$ and is most useful if instead
of $H^{3}$ model we would use the Poincare sphere model.That is instead of an
open disc $D^{2}$ model with $S_{\infty}^{1}$ being a ''circle at
infinity'',we would use an open ball $B^{3}$ model with $S_{\infty}^{2}$ being
a ''sphere at infinity''.The hyperbolic transformations in $D^{2}$ have fixed
points on $S_{\infty}^{1}$ .The hyperbolic transformations in $B^{3}$ have
fixed points on $S_{\infty}^{2}.$ The hyperbolic polygon in $H$ (or $D^{2}$)
is being replaced now with the hyperbolic polyhedron in $H^{3}$ or $B^{3}.$

\textbf{Definition} \textbf{A.2}.\textit{A Fuchsian group }$\Gamma$\textit{
is\ finitely generated if the corresponding hyperbolic polygon in }$H$\textit{
(or }$D^{2})$\textit{ has finite hyperbolic area.}

\textbf{Definition} \textbf{A.3}.\textit{A quasi-Fuchsian (Kleinian) group is
geometrically finite if the corresponding hyperbolic polyhedron in }$H^{3}%
($\textit{or }$B^{3})$\textit{ has finitely many faces.}

\textbf{Remark} \textbf{A.4}.\textit{All known knots /links (except torus and
satellite) have as their complements the hyperbolic manifolds associated with
the geometrically finite Kleinian groups}$[36].$\textit{ \quad}

Let $G\in\Gamma$ be some element of geometrically finite Kleinian
group.Consider the hyperbolic distance $d_{G}(x)=d(x,G(x)).$ Thurston$[36]$
had proven the following

\textbf{Theorem} \textbf{A.5.} $G$\textit{ is hyperbolic if and only if the
infimum of }$d_{G}$\textit{ is positive.This infimum is attained along the
line,which is unique axis for }$G$\textit{ .The endpoints of the axis are the
fixed points of }$G$\textit{ on }$S_{\infty.}^{2}$

\textbf{Corollary} \textbf{A.6}.\textit{If }$x\in H^{3}($\textit{ or }$B^{3}%
)$\textit{,the limit set }$L_{G}\subset S_{\infty}^{2}$\textit{ is the set of
accumulation points of the orbit }$G_{x}$\textit{ of x. }$L_{G\text{ }}%
$\textit{is independent of the choice of }$x.$

\textbf{Definition} \textbf{A.7.}\textit{The domain of discontinuity
}$D_{\Gamma}$\textit{ for a discrete group }$\Gamma$\textit{ is given by
}$D_{\Gamma}=S_{\infty}^{2}\backslash L_{\Gamma}.$\textit{ A discrete subgroup
of PSL(2,C) acting on }$D_{\Gamma}$\textit{ whose domain of discontinuity is
non-empty is called Kleinian group.}

\textbf{Definition} \textbf{A.8.}\textit{A group }$\Gamma$\textit{ acting on a
locally compact space X is called properly discontinuous if for every compact
set K}$\in$\textit{X there are only finitely many }$\gamma\in\Gamma$\textit{
such that }$\gamma$\textit{K}$\cap$\textit{K}$\neq\emptyset.$

\textbf{Definition A.9}.\textit{The Kleinian manifold }$\bar{M}$\textit{ is
defined by}
\[
\bar{M}=\left(  H^{3}\cup D_{\Gamma}\right)  /\Gamma\text{ .}%
\]
\quad\quad

Let $S$ be compact oriented surface of negative Euler characteristic.Let
$\partial S$ denote the boundary of $S$ and $intS=S\backslash\partial S$ be
the interior of $S$ .Analogously,for 3-manifolds we have

\textbf{Definition A.10.}\textit{The boundary of }$\bar{M}$\textit{ is given
by }$\partial\bar{M}=D_{\Gamma}/\Gamma$ .

\textbf{Corollary} \textbf{A .11}.\textit{Some Kleinian manifolds are bounded
by a pair of Riemann surfaces X(bottom) and Y(top) .There is a homeomorphism
between }$\bar{M}$\textit{ and }$intS\times\lbrack0,1]$\textit{ which is
compatible with marking of X by }$S$\textit{ and Y by }$\bar{S}$\textit{
(where }$\bar{S}$\textit{ is the same surface as }$S$\textit{ but \quad with
reversed orientation).}

\textbf{Proof}. Please,consult Refs.[70,71] and [94].$\square$

\textbf{Remark} \textbf{A.12.}\textit{The marking of Riemann surface X should
be understood in the following sense. The Teichm\"{u}ller space }%
$Teich(S)$\textit{ classifies the conformal structures on }$intS$\textit{ in
which each boundary component corresponds to a puncture.A point in }%
$Teich(S)$\textit{ is specified by X which is conformaly equivalent to a
punctured disc }$D^{2}$\textit{(surely ,in general,with more than one
puncture).A homeomorphism }$f:intS\rightarrow$\textit{X which sends the
orientation on }$S$\textit{ to the (canonical) orientation on X is called
marking of X.}

The Corollary A.11. leads us to the most important generalization due to
Bers,Ref.[71].It can be formulated in the form of the following

\textbf{Theorem} \textbf{A.13.}\textit{Given two (in general,different)Riemann
surfaces X and Y of genus g and an orientation reversing map J between
them,there is a uniquely determined (up to normalization)quasi-Fuchsian group
}$G$\textit{ for which the bottom surface is (conformaly equivalent to)X,the
top ,to Y,and the associated involution X}$\rightarrow$\textit{Y is homotopic
to J.}

\textbf{Definition A.14.}Consider the set of all quasi-Fuchsian groups
whose\textbf{\ bottom} surface X is\textbf{\ fixed }.This set is called the
\textbf{\ Bers slice }$B_{Y}$ \textbf{based on} X .The points in the
Teichm\"{u}ller space are represented by the top surfaces and their relation
(via involution) to the \textbf{fixed }bottom surface X.

\textbf{Corollary} \textbf{A.15}.\textit{The motion through the Bers slice is
associated with the motion through various hyperbolic 3-manifolds whose bottom
boundary component is fixed but whose top varies. The slice may have an edge
associated with the fact that the top component may eventually degenerate in
some way.}

\textbf{Remark} \textbf{A.16}\textit{.\textbf{a})The motion through the Bers
slice is directly connected with the motion due to the pseudo-Anosov surface
homeomorphisms(Theorem6.20)\textbf{; A.16.b})The degeneration of the Bers
slice is associated with the boundary of the Teichmuller space and formation
of all kinds of spikes(cusps) on the top surface.}

\textbf{Definition} \textbf{A.17}.\textit{3-manifold is simply degenerate if
it is topologically equivalent (homeomorphic )to the product }$S\times
R$\textit{ where }$S$\textit{ is the compact Riemann surface of negative Euler
characteristic}$[74].$

\textbf{Definition A.18.}3-manifold is called \textbf{geometrically tame} if
all of its ends are either geometrically finite(Definition A.3) or simply
degenerate (Definition A.17).

\textbf{Definition} \textbf{A.19.}\textit{The thick-thin decomposition of a
hyperbolic manifold }$M$\textit{ ,Ref.[36], can be best understood by
analyzing Fig.6.The thin part of }$M$\textit{ is made of neigbourhoods of
short geodesics and cusps isomorphic to pseudospheres}\marginpar{Fig.6.}

\textbf{Theorem} \textbf{A.20.}\textit{The complete hyperbolic manifold }%
$M$\textit{ has the finite hyperbolic volume if and only if the thick part
,}$M_{\geq\varepsilon}$\textit{ ,is compact for all }$\varepsilon>0.$

\textbf{Proof}. Please,consult Ref.36.$\square$

\textbf{Remark} \textbf{A.21.}\textit{This theorem is central for the whole
chain of arguments which provide justification of knots/links existence in 2+1
gravity.Indeed,guided by this theorem,it is possible now to prove the following}

\textbf{Theorem} \textbf{A.22}.\textit{Let }$M=H^{3}/\Gamma$\textit{ be
infinite volume topologically tame hyperbolic 3-manifold.Then,if }$\hat
{\Gamma}$\textit{ is finitely generated subgroup of }$\Gamma$\textit{ ,either}

\textit{\quad a) }$\hat{M}=H^{3}/\hat{\Gamma}$\textit{ is geometrically finite,or}

\textit{\quad b) the thin part }$M_{\leq\varepsilon}$\textit{ has a
geometrically infinite end }$\hat{E}$\textit{ such that the local}

\textit{\quad isometry }$p:\hat{M}\rightarrow M$\textit{ is finite one-to-one
on some neigbourhood }$\hat{U}$\textit{ of }$\hat{E}.$

\textbf{Proof}. Please,consult Ref.[96] and take into account Fig.6 and
Theorem A.20.$\square$

\textbf{Remark} \textbf{A.23} \textit{Although the TheoremA.22. provides the
desired result ,it does not contain the constructive prescription as to how to
obtain }$\hat{M}$\textit{ (that is how to find }$\hat{\Gamma}$\textit{ and to
construct the quotient).The most spectacular results of Ref.70 address just
this problem.They are the following:}

\quad a)\textit{ the manifold }$\hat{M}$\textit{ is homeomorphic to the
hyperbolic 3-manifold which fiber}

\textit{\quad\quad over the circle;}

\textit{\quad b) the motion through the Bers slice which produces }$\hat{M}%
$\textit{ is characterized}

\textit{\quad\quad by the functional equation ( similar to the Feigenbaum-Cvitanovich}

\textit{\quad\quad equation for the chaotic maps}$[97])$\textit{ which
possesses the fixed point}

\textit{\quad\quad controlling the existence of }$\hat{M}$\textit{ .}

\textbf{Remark} \textbf{A.24}.\textit{In view of the Feigenbaum-Cvitanovich
universal equation for maps of the interval ,it is reasonable to expect some
sort of universality in the knot /link structures (types) for 2+1 gravity.This
topic deserves much more study.}

\textit{\mathstrut}

\newpage

\mathstrut

\mathstrut

\quad\quad\quad\quad\quad\quad\quad\quad\quad\quad\textbf{Figure Captions}

\mathstrut Fig.1 Conformal mapping between the rectangle and the annulus

\mathstrut

\bigskip Fig.2 Conformal mapping between w and z planes which explains how

Y-type singularity can be built out of ''flat bricks''(rubber bands)

\bigskip

Fig.3 Conformal mapping between w and z planes which explains

how thorn can be built from the''flat brick''(rubber band)

\bigskip

Fig.4 Some of the ''forbidden'' sigularities

\mathstrut

Fig.5 The topological structure of every closed Riemann surface R

is encoded in 3-valent planar graph G.With help of this graph the pants

decomposition can be found.The graph G provides the gluing prescription

for such pantsdecomposition

\medskip

Fig.6 The thin-thick decomposition of a hyperbolic surface

\newpage

\mathstrut

\mathstrut

\quad\quad\quad\quad\quad\quad\quad\quad\quad\quad\quad\quad\quad\quad\textbf{References}

[1] A.Kholodenko,Use of meanders and train tracks for description of

defects and textures in liquid crystals and 2+1 gravity (Part I).

[2] J.Hubbard and H.Masur ,Quadratic differentials and foliations,

Acta Math.142(1979)221 -274.

[3]V.Poenaru , Some aspects of the theory of defects in ordered

media and gauge fields related to foliations,Comm.Math.Physics

80\textbf{(}1981\textbf{)}127-136.

[4] R.Langevin ,Foliations,energies and liquid crystals ,

Asterisque 107-108(1983)201-213.

[5] B.Zweibach ,How covariant closed string theory solves the

minimal area prolem ,Comm.Math.Physics 136(1991)83-118.

[6] M.Kontsevich ,Intersection theory on the moduli space of curves

and the matrix Airy function, Comm.Math.Physics147(1992)1-23 ).

[7] \ K.Strebel , \textit{Quadratic Differentials },Springer-Verlag, Berlin,1984.

[8]F.Gardiner,\textit{Teicm\"{u}ller Theory and Quadratic Differentials ,}

Wiley Interscience, New York,1987.

[9] S.Deser , R.Jackiw and G. 't Hooft , Three dimensional Einstein

gravity :dynamics of flat space ,Ann.Phys.152\textbf{ (}1984\textbf{),}220-235 .

[10].M.Troyanov , Les surfaces euclidiennes a singularities coniques,

L'Enseignement Math.32(1986)79-94.

[11] I.Rivin ,Euclidean structures on simplicial surfaces and hyperbolic

volume, Ann.Math.139(1994) 553-580.

[12].H.Hopf , \textit{Differential Geometry in the Large} ,LNM No.1000

,Springer-Verlag,Berlin,1989.

[13]H.Masur and J.Smillie,Quadratic differentials with prescribed

singularities and pseudo-Anosov diffeomorphisms ,Comm.Math.

Helvetici\textbf{\ }68\textbf{ }(1993)289 -307.

[14] L.Takhtadjian ,\textit{Topics in Quantum Geometry of Riemann Surfaces:Two}

\textit{Dimensional Quantum Gravity,}hep-th /9409088.

[15] V.Moncrief ,Reduction of the Einstein equations in 2+1 dimensions

to a Hamiltonian system over Teichm\"{u}ller space,

J.Math.Phys.\ 30(1989)\textbf{ },2907 -2914.

[16] A.Kholodenko and T.Vilgis , Some Geometrical and Topological

Problems in Polymer Physics, Physics Reports 298\textbf{(}1998\textbf{)}251-372.

[17].A.Kholodenko ,Statistical mechanics of the deformable droplets

on flat surfaces J.Math.Phys. 37\textbf{ }(1996)1287-1313;ibid Statistical

mechanics of the deformable droplets on Riemann surfaces:applications

to reptation and related problems ,1314-1335 .

[18] D.Mitrinovich , \textit{Analytic Inequalities} ,Springer-Verlag ,Berlin,1970.

[19] T.Rado, \textit{Lenght and Area} , Am.Math.Soc.Colloq.Publ.30

,Am.Math.Soc.,Providence,RI,1948.

[20] R.Osserman ,The isoperimetric inequality, Bull.Am.Math.

Soc. 84\textbf{ }(1978)1182-1238 .

[21] L.Ahlfors , \textit{Lectures on Quasiconformal Mappings }

,D.van Nostrand Co.Princeton,1966.

[22] M.Reed and B.Simon , \textit{Methods of Modern Mathematical Physics,}Vol.1

,Academic press,New York, 1972.

[23] S.Chandrasekhar , \textit{Liquid Crystals} ,Cambridge U.Press,

Cambridge,1992 .

[24] G.Levitt ,Foliations and laminations on hyperbolic surfaces ,

Topology 22 (1983)119-135 .

[25] K.Strebel , On quadratic differentials with closed trajectories

and second order poles, J.Analyse Math.19\textbf{ }(1983)373-382 .

[26] K.Strebel , On quadratic differentials with closed trajectories

on open Riemann surfaces, Ann.Acad.Sci.Fenn.A2\textbf{ }(1976) 533 -551.

[27] P.Buser,\textit{\ Geometry and Spectra of Compact Riemann Surfaces}

,Birkh\"{a}user,Boston,1992 .

[28] A.Marden , Geometric relations between homeomorphic

Riemann surfaces ,Bull.Am.Math.Soc.(New Series)3\textbf{ }(1980)1001-1017 .

[29] W.Abikoff ,\textit{\ The Real Analytic Theory of Teichm\"{u}ller Space}

,Springer-Verlag , New York,1980.

[30] S.Chern in \textit{Analysis Et Cetera },Academic Press,New York ,1980.

[31] W.Fulton , \textit{Algebraic topology.A first Course}

,Springer -Verlag,Berlin,1995.

[32] A.Casson and S.Bleiler , \textit{Automorphisms of Surfaces After }

\textit{Nielsen and Thurston }, Cambridge U.Press,Cambridge,1993.

[33]H.Poincare,\textit{\ Collected Works} ,Nauka,Moscow,1971.

[34] A.Fathi , F.Laudenbach and V.Poenaru ,Travaux de Thurston

sur les diffeomorphismes des surfaces et l'espace de Teichm\"{u}ller,

Asterisque 66-67 (1979).

[35].J-P. Otal ,Le theoreme d'hyperbolization pour les varietes fibres

de dimension 3, Asterisque 235\textbf{ }(1996).

[36] W.Thurston , \textit{Three -Dimensional Geometry and Topology} ,Vol.1

,Priceton U.Press,Princeton,1997.

[37] P.-G.de Gennes ,\textit{\ The Physics of Liquid Crystals }

,Clarendon Press,Oxford, 1979.

[38] S.Singh ,Curvature elasticity in liquid crystals ,

Phys.Reports 277\textbf{ }(1996)283 -386.

[39] A.Polyakov ,\textit{\ Gauge Fields and Strings },Harwood

Academic, New York,1987.

[40] H.Bresis , J.-M. Coron and E.Lieb ,Harmonic maps with

defects ,Comm.Math.Phys. 107\textbf{ }(1986)649 -705.

[41].H.Bresis in \textit{Topics in Calculus of Variations },LNM No1365

,Springer-Verlag,Berlin, 1988.

[42] F.Bethuel ,H.Bresis and F.Helein , \textit{Ginzburg-Landau Vortices }

,Birkh\"{a}user,Boston,1994.

[43] P.Chaikin and T.Lubensky , \textit{Principles of Condensed Matter}

\textit{ Physics },Cambridge U.Press,Cambridge,1995..

[44] R.Courant , \textit{Geometrisce Function Theorie} ,

Springer-Verlag,Berlin, 1964.

[45]A.Marden and K.Strebel , The heights theorem for quadratic

differentials on Riemann surfaces ,Acta Math. 153\textbf{ }(1984)153-211 .

[46] J.Stillwell ,\textit{\ Geometry of Surfaces },Springer-Verlag,Berlin, 1992.

[47].B.Halperin and D.Nelson , Theory of two dimensional melting,

Phys.Rev.Lett. 41\textbf{ ,}121(1978).

[48] J.-B.Bost in \textit{From Number Theory to Physics} ,Springer-Verlag,

Berlin, 1992.

[49] L.Hugston and K.Tod , \textit{An Introduction to General Relativity}

,Cambridge U.Press,Cambridge,1990.

[50] R.Wald ,\textit{\ General Relativity },U.of Chicago Press ,Chicago,1984.

[51]A.Petrov , \textit{Einstein Spaces }(Pergamon,London,1969).

[52] B.Dubrovin ,A.Fomenko and S.Novikov ,\textit{\ Modern }

\textit{Geometry-Methods and Applications },Vol.1.

,Springer-Verlag,Berlin, 1984.

[53].E.Witten ,2+1 dimensional gravity as an exactly soluble problem,

Nucl.Phys. B311 (1988)46-78 .

[54] F.Luo and G.Tian ,Liouville equation and spherical complex

polytopes, Proc.Am.Math.Soc.116\textbf{(}1992\textbf{)}1119 -1129.

[55] R.Mc Owen , Point singularities and conformal metrics on

Riemann surface, Proc.of Am. Math.Soc.103\textbf{ }(1988)222--225.

[56] R.Nevanlinna ,\textit{\ Analytic Functions },Springer-Verlag,Berlin, 1965.

[57].J.Jenkins , \textit{Univalent Functions and Conformal Mapping }

,Springer-Verlag,Berlin, 1965.

[58] L.Bers ,Quasiconformal mapping with applications to differential

equations,function theory and topology, Bull of Am.Math.Soc.

83\textbf{ }(1977)1083 -1100.

[59] O. Lehto , \textit{Univalent Functions and Teichm\"{u}ller Spaces }

,Springer-Verlag,Berlin, 1987.

[60] L.Bers , An extremal problem for quasiconformal mapping and

a theorem by Thurston ,Acta Math.141\textbf{ }(1978)73-98

[61] A.Marden , and K.Strebel ,Pseudo-Anosov Teichm\"{u}ller mapping,

J.d'Analyse Math. 46\textbf{ }(1986)194 -220.

[62].L.Slutskin ,Classification of lifts of automorphisms of surfaces

to the unit disc,Contemporary Math.152\textbf{ }(1993) 311-340 ..

[63] J.Hempel, 3-\textit{Manifolds },Princeton U.Press ,Princeton,1976.

[64] K.Johannson , \textit{Topology and Combinatorics of 3-Manifolds }

,Springer-Verlag,Berlin, 1995.

[65] G.Hemion,\textit{\ The Classification of Knots and 3-Dimensional Spaces}

,Oxford U.Press.Oxford,1992).

[66] W.B.R.Lickorish, A representation of orientable combinatorial 3

manifolds, Ann.of Math. 76\textbf{ }(1962)531-540.

[67].A.Wallace ,Modifications and coboundary manifolds,

Canadian J.of Math.12\textbf{ }(1960) 503-528 (1960).

[68] R.Geroch , Domain of dependence, J.Math.Phys.

11\textbf{ (}1970\textbf{)}437-449 \textbf{.}

[69] S.Hawking and G.Ellis , \textit{The Large Scale Structure of Space-Time }

,Cambridge U.Press,Cambridge,1973.

[70]C.McMullen ,\textit{\ Renormalization and 3-Manifolds Which Fiber Over the}

\textit{Circle }(Princeton U.Press ,Princeton,1996).

[71] L.Bers, Fiber spaces over Teichm\"{u}ller spaces ,

Acta Math.130(1972)\textbf{ }89 -126.

[72] E.D'Hoker and D.Phong ,The geometry of string

perturbation theory,Rev.Mod.Phys.\textbf{\ }60\textbf{ }(1988)917- 1066.

[73] G.'t Hooft ,Causality in 2+1 dimensional gravity,

Class.Quantum Grav.9 \textbf{(}1992)1335 -1348.

[74] G.'t Hooft ,Evolution of gravitating point particles in 2+1 

dimensions, Class.Quantum Grav.10\textbf{(}1993\textbf{)}1023-1038.

[75]G.'t Hooft ,Canonical quantization of gravitating point particles 

in 2+1 dimensions,Class.Quantum Grav.10(1993) 1653-1664.

[76]G.'t Hooft ,Quantization of point particles in 2+1 dimensions

and spacetime discretness,Class.Quantum Grav.13\textbf{ (}1996\textbf{)}1023 -1039.

[77] R.Franzosi and E.Guadagnini ,Topology and classical geometry

in 2+1 gravity,Class.Quantum Grav.13 (1996)433 -460.

[78] A.Connes , D.Sullivan and N.Teleman ,Quasiconformal mappings,

operators on Hilbert space and local formulas for characteristic classes,

Topology 33\textbf{ }(1994) 663-684.

[79] W.Veech ,The Teichm\"{u}ller geodesic flow,Ann.Math.

124\textbf{ }(1986) 441-530

[80] S.Kerkhoff , H.Masur and J.Smillie ,\ Ergodicity of billiard flows

and quadratic differentials, Ann.Math.124\textbf{ (}1986\textbf{)}293 --311..

[81]Y.Vorobets, Plane structures and billiards in rational polygons:

the Veech alternative ,Russian Math.Surveys 51\textbf{ }(1996)779-817.

[82] H.Masur , Closed trajectories for quadratic differentials with an

application to billiards, Duke Math.Journal 53\textbf{ }(1986)307-314.

[83] H.Masur ,Interval exchange transformations and measured foliations,

Ann.Math. 115 \textbf{(}1982\textbf{)}169 -200.

[84].J.Math.Phys.37(10) (1996).Special Issue on Mesoscopic Systems

[85] A.Lobkovsky and T.Witten ,Properties of ridges in elastic

membranes, Phys.Rev.E 55\textbf{ }(1977)1577-1599.

[86] A.Lobkovsky, S.Gentges ,H.li ,D.Morse and T.Witten,

Scaling properties of stretching ridges in a crumpled elastic sheet,

Science 270\textbf{ }(1995)1482-1485.

[87] K.Leung and J.Andersen, Phase transition in aspring-block model 

of surface fracture,Europhys.Lett. 38(1997)589 -594.

[88] O.Huseby , J.-F. Thovert and P.Adler ,Geometry and topology

of fracture systems, J.Phys.A30 (1997)1415-1444.

[89] A.Buchel and J.Setna , Statistical mechanics of cracks:

Fluctuations,breakdown,and asymptotics of elastic theory,

Phys.Rev.E 55\textbf{ (1997)}7669 -7690.

[90]W.B.R.Lickorish ,Invariants for 3-manifolds from the

combinatorics of the Jones polynomial, Pac.Journ.of Math.

149 (1991)337-347.

[91] W.B.R.Lickorish ,Three-manifolds and the Temperley-Lieb

algebra, Math.Ann. 290\textbf{ }(1991)657-670.

[92] E.Witten , Quantum field theory and the Jones polynomial,

Comm.Math.Phys. 121\textbf{ }(1988)351-399..

[93] B.Westbury , The representation theory for the

Temperley-Lieb algebra, Math.Zeitschrift 219\textbf{ }(1995)539 -565.

[94] A.Marden , Geometric relations between homeomorphic

Riemann surfaces,Bull.Am.Math.Soc.(New Series) 3\textbf{ }(1980)1001-1017.

[95] Y.Minsky ,Teichmuller geodesics and ends of hyperbolic 3-manifolds,

Topology ,32(1993) 625-647.

[96] R.Canary , A covering theorem for hyperbolic 3-manifolds and its

applications, Topology 35\textbf{ }(1996)751-778.

[97] P.Cvitanovich ,\textit{\ Universality in Chaos },Adam Hilger Ltd ,Bristol,1984.
\end{document}